\documentclass[twocolumn,superscriptaddress,showpacs,pra, aps, 10pt]{revtex4-1}

\usepackage{amssymb,bm, amsbsy,
	amsmath, amsthm, latexsym, dsfont, array,subfigure, layout, graphicx, mathrsfs,color,soul}
\usepackage[colorlinks=true,citecolor=blue,linkcolor=blue,anchorcolor=blue,urlcolor=blue]{hyperref}
\hypersetup{linktoc=all}

\allowdisplaybreaks
\newcommand{\ket}[1]{\left|{#1}\right\rangle}
\newcommand{\bra}[1]{\left\langle{#1}\right|}

\newcommand{\ketbra}[2]{\left|{#1}\rangle\!\langle{#2}\right|}

\renewcommand{\l}{\left}

\renewcommand{\d}{\rm d}
\renewcommand{\r}{\right}
\newcommand{\E}{\mathcal{E}}

\renewcommand{\url}[1]{}
\newcommand{\urlprefix}{}
\renewcommand{\href}[1]{}
\newcommand{\blk}{\color{black}}

\newcommand{\tr}{{\rm Tr}}
\newcommand{\rhos}{\rho_{\rm S}}
\newcommand{\rhob}{\rho_{\rm B}}

\usepackage{xcolor}
\definecolor{CHIGUSA}{RGB}{58,143,183}
\definecolor{KON}{RGB}{193,105,60}
\definecolor{HATOBANEZUMI}{RGB}{114,99,110}

\begin{document}
	
	\title{Boson-boson pure-dephasing model with non-Markovian properties}
	\author{Fei-Lei Xiong}
	\affiliation{Hefei National Laboratory for Physical Sciences at the Microscale, Department of Modern Physics, University of Science and Technology of China, Hefei, Anhui 230026, China}
	\affiliation{CAS Center for Excellence in QIQP and the Synergetic Innovation Center for QIQP, University of Science and Technology of China, Hefei, Anhui 230026, China}
	\author{Li Li}
	\email{kenny.li@griffithuni.edu.au}
	\affiliation{Centre for Quantum Computation and Communication Technology (Australian Research Council), Centre for Quantum Dynamics, Griffith University, Brisbane, Queensland 4111, Australia}
	\author{Zeng-Bing Chen}
	\affiliation{Hefei National Laboratory for Physical Sciences at the Microscale, Department of Modern Physics, University of Science and Technology of China, Hefei, Anhui 230026, China}
	\affiliation{CAS Center for Excellence in QIQP and the Synergetic Innovation Center for QIQP, University of Science and Technology of China, Hefei, Anhui 230026, China}
	
\begin{abstract}
Understanding decoherence processes is crucial in the study of open quantum systems. In this paper, we discuss the mechanism of pure-dephasing process with a newly proposed boson-boson model, namely, a bosonic field coupled to another bosonic bath in thermal  equilibrium. Our model is fully solvable and can reproduce the pure-dephasing process which is usually described by the well-known spin-boson model, therefore offering a new perspective to understanding decoherence processes in open quantum systems of high dimension. 
We also show that this model admits a  generically non-Markovian dynamics with respect to various different non-Markovian measures. 
\end{abstract}
	
	
\maketitle
	
\section{INTRODUCTION}
\label{sec-intr}
	
Lying at the heart of many intriguing phenomena of quantum systems, such as quantum interference~\cite{ANV09} and quantum entanglement~\cite{HHH09}, superposition property plays a fundamental role in quantum theory.
However, in reality, quantum systems are usually open in the sense that they are inevitably coupled to other systems (usually referred to as the environment/reservoir/bath). The superposition property of an open quantum system is gradually lost during its dynamics, a process well known as \emph{decoherence}~\cite{Z03}. In some cases, if the decay rates of the non-diagonal elements of the system density matrix are much faster than the rate at which the system approaches equilibrium, i.e., the diagonal terms are approximately constants within that time scale, those dynamics usually can be modeled as pure-dephasing processes~\cite{GZ04}. 
As an important class of decoherence processes, understanding pure-dephasing processes is of great importance both in applications and concepts. They have been widely used for studying photon emission~\cite{AGP09,NSK08} and quantum information processing~\cite{PSH08,U95,PSE96,DG98,RQJ02}. From the fundamental perspective, they are also useful for characterizing quantum phase transition~\cite{QSL06}, quantum Zeno (anti-Zeno) effects~\cite{CG14}, quantum-classical transition~\cite{Z82}, and memory effects in open quantum systems~\cite{ZTZ11,HJM13,ABH14,ABC14,ALT15}. 

Roughly, pure-dephasing processes fall into two categories: spin-spin interactions~\cite{CPZ05,RCG07,RCG072,HGM12} and spin-boson interactions~\cite{GJC10,YL08,LWS08}. A distinct application of the former case is that the system (a spin) could act as a probe of quantum phase transition~\cite{QSL06,HGM12}, and the configuration can be implemented with atoms in optical lattices~\cite{RCG07,RCG072}. While for the latter case, it plays an important role in designing quantum registers~\cite{RQJ02,DG98}, which is a critical step toward manufacturing quantum computers~\cite{U95,PSE96}.

In this paper, we propose another pure-dephasing model, i.e., a boson-boson pure-dephasing model. In this model, the system of interest is a single-mode bosonic field; the bath is composed of many bosonic fields; and the system interacts with the bath modes through cross-Kerr interaction. In the literature, the cross-Kerr interaction between two bosonic modes (usually photonic modes) is widely studied for the implementation of quantum gates~\cite{CY95} and the preparation of entangled states~\cite{DGCZ00,PKH03}. The situation studied here is that many modes (the bath) interacts with a mode (the system) through the cross-Kerr interaction.
Because the interaction term is the product of the system Hamiltonian and an operator of the bath, the dynamics is pure dephasing.

Charactering non-Markovianity in open quantum systems has received great attention in recent years~\cite{BLP09,RHP10}. In fact, this property has been widely studied in spin-spin~\cite{CPZ05,RCG07,RCG072,HGM12} and spin-boson models~\cite{GJC10,YL08,LWS08}. In this paper, we show that the boson-boson interaction also admits non-Markovian dynamics for the system. However, as shown in Ref.~\cite{LHW17}, since there is still no consensus on the definition of quantum non-Markovianity~\cite{MJL14}, we employ three different non-Markovian criteria, namely, divisibility~\cite{RHP10}, {quantum regression formula}~\cite{GSV14}, and the Wigner function~\cite{WM07,BV05}. 

The paper is structured as follows. In Sec.~\ref{sec-pure-dep}, we propose the boson-boson pure-dephasing model, and give the exact solution of the system state when the initial bath state is a thermal state. We show that the coherent property, i.e., the off-diagonal elements of the system density matrix, can be characterized by one particular function.  In Sec.~\ref{sec-rep-sb}, we show that our model can reproduce the pure-dephasing of a spin. Then we compare the dynamics with that of a spin-boson pure-dephasing model.  In Sec.~\ref{sec-non-Mar}, we analyze the non-Markovian property of the model. Finally, we summarize the main results in Sec.~\ref{sec-con}.

\section{The Boson-Boson PURE-DEPHASING MODEL}
\label{sec-pure-dep}
	
Consider an open quantum system consisting of a single-mode bosonic field which interacts with a bosonic bath. Following the standard description of open quantum systems~\cite{BP02}, the total Hamiltonian can be divided into three parts
\begin{align}
H_{\rm tot} = H_{\rm S}+H_{\rm B}+H_{\rm SB}\, ,
\end{align}
where $H_{\rm S}$, $H_{\rm B}$, and $H_{\rm SB}$ represents  the Hamiltonian of the system, the bath and the interaction respectively. For the boson-boson pure-dephasing model, they can be explicitly written as ($\hbar\equiv 1$, $k_{\rm B}\equiv 1$ throughout the text)
\begin{align}
H_{\rm S} &=\omega_{\rm S} b^{\dagger}b \, , \label{eq-hs} \\
H_{\rm B} &=\sum _k \omega _k  b_k^{\dagger } b_k \, , \label{eq-he} \\
H_{\rm SB} &=\sum _k \lambda_k b^{\dagger} b b_k^{\dagger} b_k  \, , \label{eq-Hinter}
\end{align}
where $\omega_{\rm S}$, $b^{\dagger}$, and $b$ are the single-particle energy, creation operator, and annihilation operator of the system, respectively; $\omega_k$, $b^{\dagger}_k$, and $b_k$ are the single particle energy, creation operator, and annihilation operator of the $k$th mode of the bath, respectively; $\lambda_k$ is the coupling strength. It is worth pointing out that each component, $\lambda_k b^{\dagger} b b_k^{\dagger} b_k $, in $H_{\rm SB}$ represents the cross-Kerr interaction in quantum optics~\cite{SZ99}, which indicates a possible physical implementation of the boson-boson model.


From the Hamiltonian in Eqs.~\eqref{eq-hs}-\eqref{eq-Hinter}, it is easy to see that $H_{\rm S}$ commutes with $H_{\rm SB}$. That is, the particle number (hence the energy) of the system is conserved during the dynamics. In fact, the system will undergo a pure-dephasing process in the energy basis due to the interaction, which will be discussed in detail in following subsections.\blk

\subsection{Exact solution of the system state}
In the Schr\"{o}dinger picture, the system state can generally be written as
\begin{align}
\label{eq-reduceS-state}
\rho_{\rm S} (t)=\tr_{\rm B}{\{U(t)\rho_{\rm SB}(0)U^{\dagger}(t)\}}\, ,
\end{align}
where $\rho_{\rm SB}(0)$ is the state of the total system at the initial time and $U(t)=\exp\{-{i}(H_{\rm S}+H_{\rm B}+H_{\rm SB})t\}$.
Now we show that our model admits an exact solution of $\rhos(t)$.
Assume that the initial state is a product state, i.e., $\rho_{\rm SB}(0)=\rhos(0)\otimes \rhob$, where $\rhos(0)$ and $\rhob$ are the initial system and bath state respectively.
One can rewrite $\rhos(0)$ in the Fock state basis as $\rho_{\rm S}(0)=\sum_{m,n}C_{mn}|m\rangle \langle n|$. It then follows from Eq.~\eqref{eq-reduceS-state} that
	\begin{align}\label{eq-densitymatrix}
	\rho_{\rm S} (t)=\sum_{m,n} C_{mn} e^{-{i}(m-n) \omega_{\rm S} t} e^{\eta((m-n)t)} |m\rangle \langle n|\, ,
	\end{align}
	where
	\begin{align}
	\label{eq-densitymatrix-coe}
	\begin{split}
	e^{\eta((m-n)t)}=\operatorname{Tr_{\rm B}}\{e^{-{i}(m-n)t \sum_k {\lambda_k \hat{n}_k}}\rho_{\rm B}\}\, .
	\end{split}
	\end{align}
and $\hat{n}_k=b_k^{\dagger} b_k$ is the particle number operator of the $k$th mode of the bath.

For any time $t$, it is easy to see from Eqs.~\eqref{eq-densitymatrix} and~\eqref{eq-densitymatrix-coe}  that the diagonal matrix element $C_{nn}(t)=C_{nn}$, i.e., the distribution of the particle number of the system is invariant in time. Decoherence manifests itself in off-diagonal terms in Eq.~\eqref{eq-densitymatrix}.  This means that our model describes a pure-dephasing process which is characterized by the factors $e^{\eta((m-n)t)}$ ($m \neq n$).  We call $\eta(t)$ the dephasing function and $e^{\eta(t)}$ the dephasing factor, which, as we will show,  fully account for the decoherence process in our model. We further assume that the bath at the initial time is in a thermal state, i.e., $\rho_{\rm B}=e ^{-H_{\rm B}/T}/Z$, where $T$ is the temperature of the bath and $Z=\operatorname{Tr}_{\rm B}\{e ^{-H_{\rm B}/T}\}$ is the bath's partition function.  In this case, the dephasing function $\eta(t)$ can be explicitly written as
	\begin{align}
	\eta(t) =\sum_k {\ln\left(\frac{1-e^{- {\omega_k }/{T}}}{1-e^{-{\omega_k }/{T}-{i}\lambda_k t}}\right)}\, .
	\end{align}
In the continuous limit of the bath modes, $\eta(t)$ reads
\begin{align}
\label{eq-zeta}
{\eta} (t)= \int_{0}^{\infty} {\mathrm{d}\omega g_{\omega} \ln\left(\frac{1-e^{- \omega/{T}}}{1-e^{- \omega/{T}-{i}\lambda_{\omega} t}}\right)}\, ,
\end{align}
where the subscript $\omega$ stands for the $\omega$-frequency mode in the bath; $g_{\omega}$ is the density of states of bath; and $\lambda_{\omega}$ is the coupling strength function. The real part of ${\eta} (t)$ characterizes the time-dependence of the moduli of the off-diagonal density matrix elements, while the imaginary part of $\eta(t)$ accounts for the system energy shift. \blk Furthermore, whether the dynamics is Markovian or not depends on both $\mathrm{Re}({\eta} (t))$ and $\mathrm{Im}({\eta} (t))$ (see Sec.~\ref{sec-non-Mar}). 

Note that $\eta (t)$ in Eq.~\eqref{eq-zeta} is the integral of $\ln\left(\frac{1-e^{- \omega/{T}}}{1-e^{- \omega/{T}-{i}\lambda_{\omega} t}}\right)$, with weight $g_\omega$. This $\log$ function is periodic with period $2\pi/\lambda_\omega$ and the amplitude  depends on the temperature $T$. In fact, as the temperature increases, the thermal fluctuation becomes more intensive, leading to stronger dephasing dynamics of the system. \blk

In the following, we will discuss the properties of $\eta(t)$ with some typical choices of the density function $g_{\omega}$ and the coupling function $\lambda_{\omega}$.

\subsection{Characterizations of the dephasing function}
\label{sec-2B}

We first choose the density of states $g_\omega$ as follows:
\begin{align}
\label{eq-gomega}
g_\omega = {\omega^s}/g\, ,
\end{align}
where $g$ is a constant~\footnote[2]{The spectral density of the bath is often denoted as $J(\omega)=A \omega^s {e}^{-\omega/\omega_c}$, where $A$ is a constant, ${e}^{-\omega/\omega_c}$ is the cutoff function and $\omega_c$ is the frequency cutoff~\cite{ZLX12}. Note that the coupling strength is also usually set as a constant. Therefore, the density of states can be written as $g_{\omega}=\omega^s {e}^{-\omega/\omega_c}/g$. However, the cutoff is unnecessary in our model, so we choose the density of states in the form of Eq.~\eqref{eq-gomega} \blk} with $s> -1$. (The constant $g$ only impacts the scale of $\eta(t)$, thus we set $g=1$ hereafter.) Then we consider three different choices of the coupling function $\lambda_\omega$. These three cases include that $\lambda_\omega$ is the same for all $\omega$; some modes are coupled more intensively with the system; and $\lambda_\omega$ is a monotonic function of $\omega$.  

First, let $\lambda_{\omega}=c$, where $c$ is a constant. Eq.~\eqref{eq-zeta} thus can be rewritten as
\begin{align}
{\eta} (t) =-\Gamma(1+s) T^{1+s} \l(\operatorname{Li}_{2+s}(1)-\operatorname{Li}_{2+s}(e^{-{i} c t}) \r)\, ,
\end{align}
where $\Gamma(1+s)$ denotes the Gamma function, and $\operatorname{Li}_{2+s}(z)=\sum_{k=1}^{\infty}{z^k}/{k^{2+s}}$ denotes the polylogarithm function~\cite{O10}. $\eta(t)$ in Eq.~\eqref{eq-zeta} has a period $2\pi/\lambda$. That is, the system state will always recur after certain time, which means there must be information backflow~\cite{BLP09}. In this sense, the system dynamics is non-Markovian.
	
\begin{figure}
\centering
\subfigure[$s=0$]{\label{fig-PlotR03}
\includegraphics[width=0.25\textwidth]{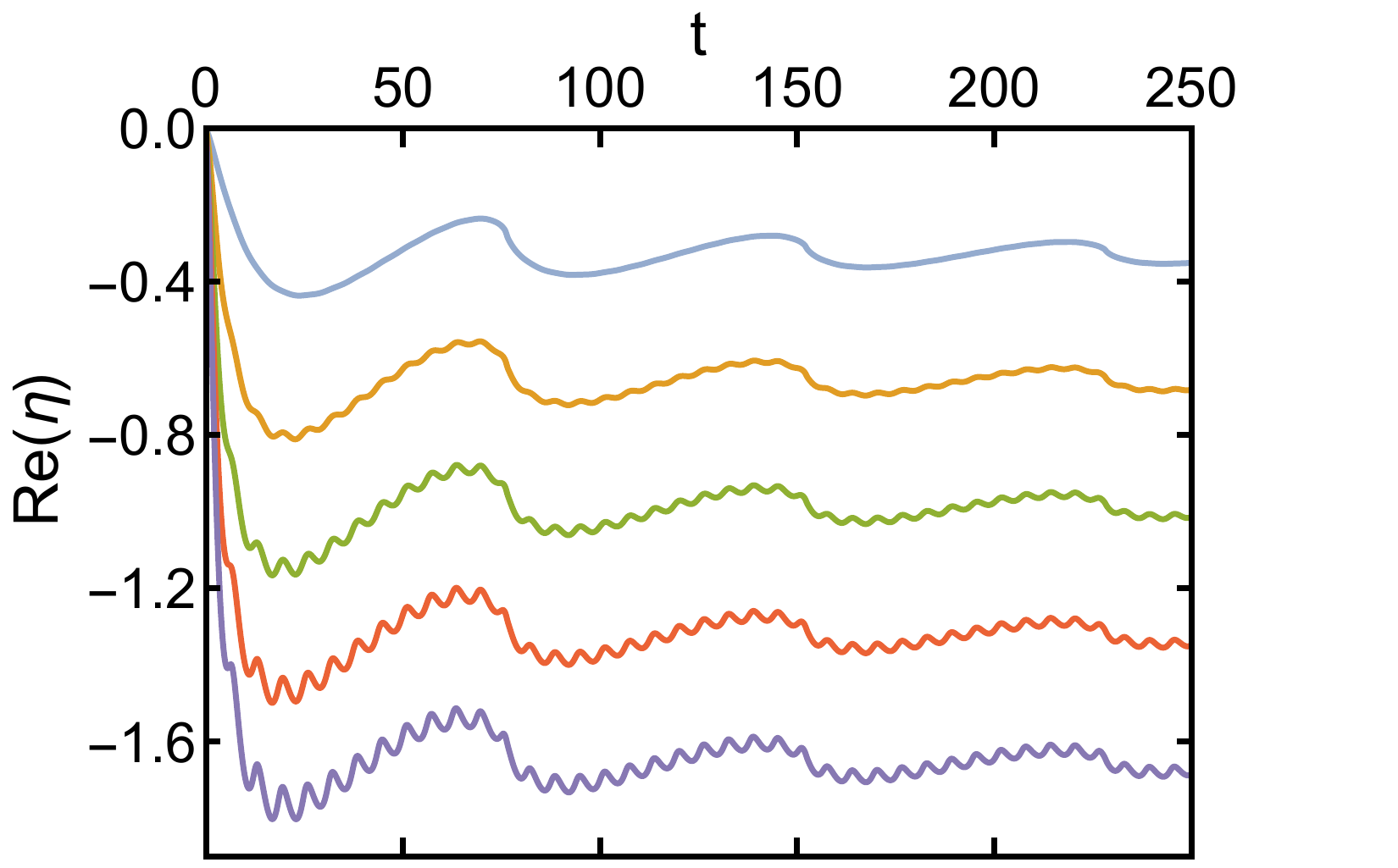}}\hspace{-2.2em}    
\subfigure[$s=0$]{\label{fig-PlotI03}
\includegraphics[width=0.25\textwidth]{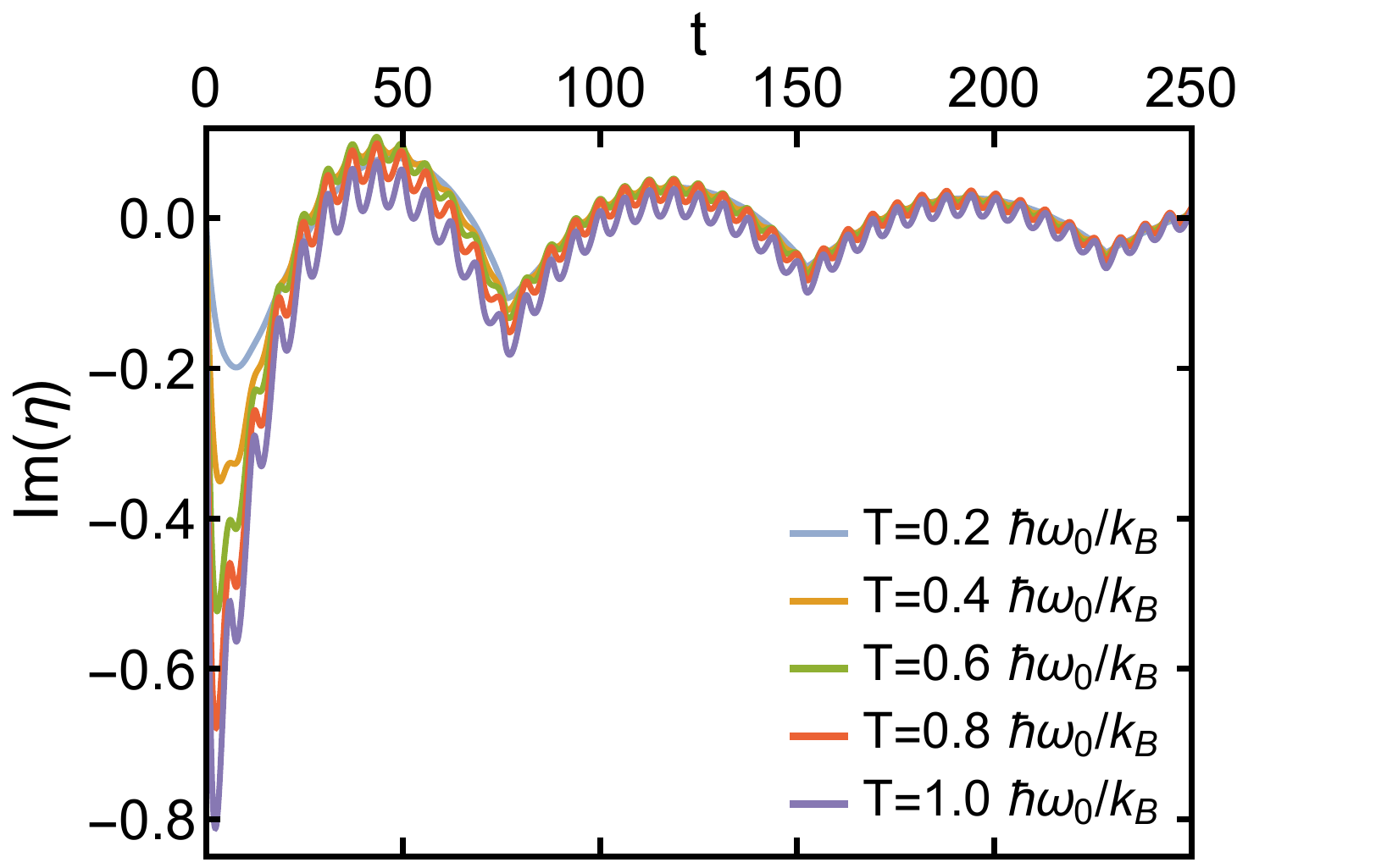}}\\[-2ex]
\subfigure[$s=1$]{\label{fig-PlotR13}
\includegraphics[width=0.25\textwidth]{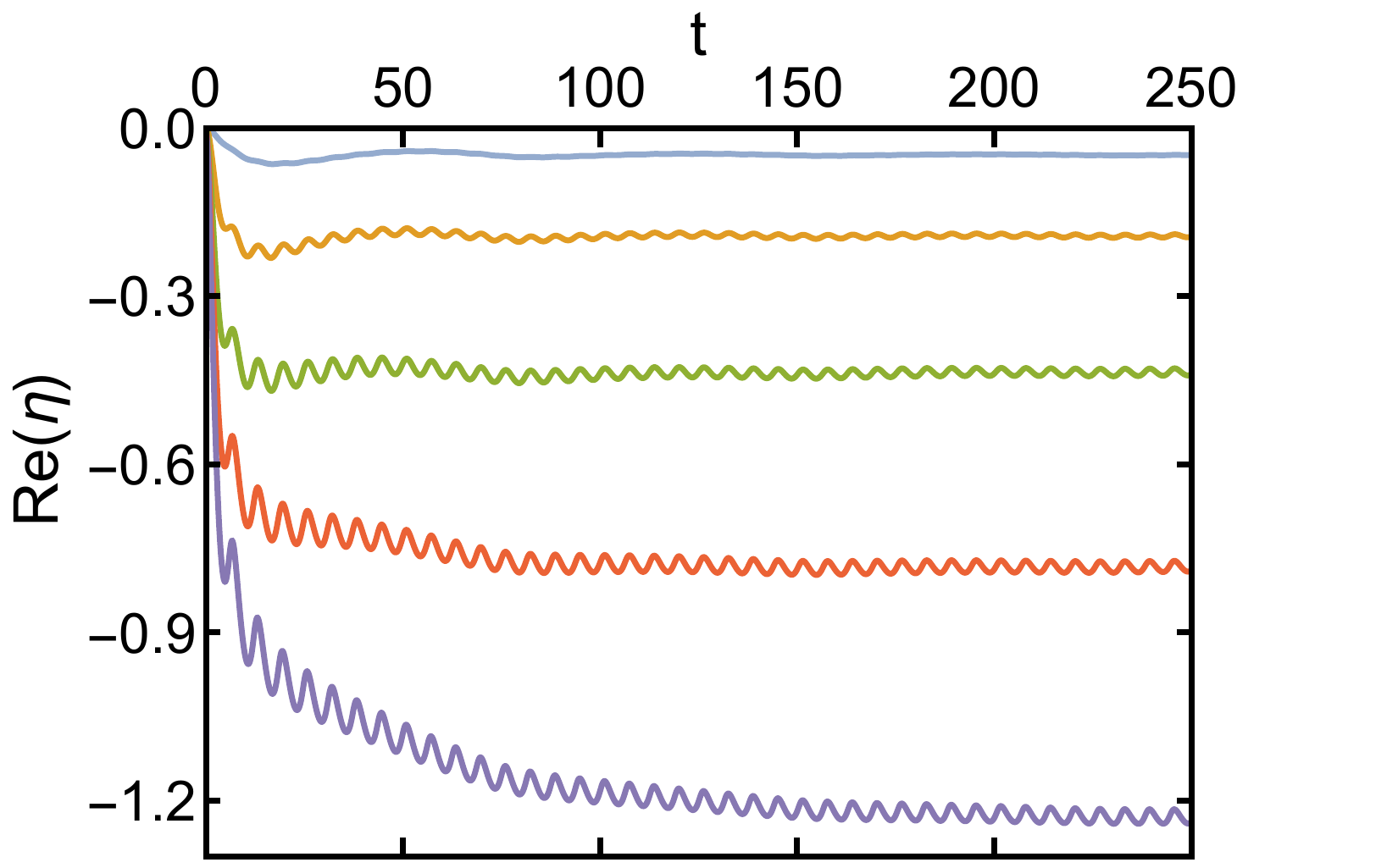}}\hspace{-2.2em}        
\subfigure[$s=1$]{\label{fig-PlotI13}
\includegraphics[width=0.25\textwidth]{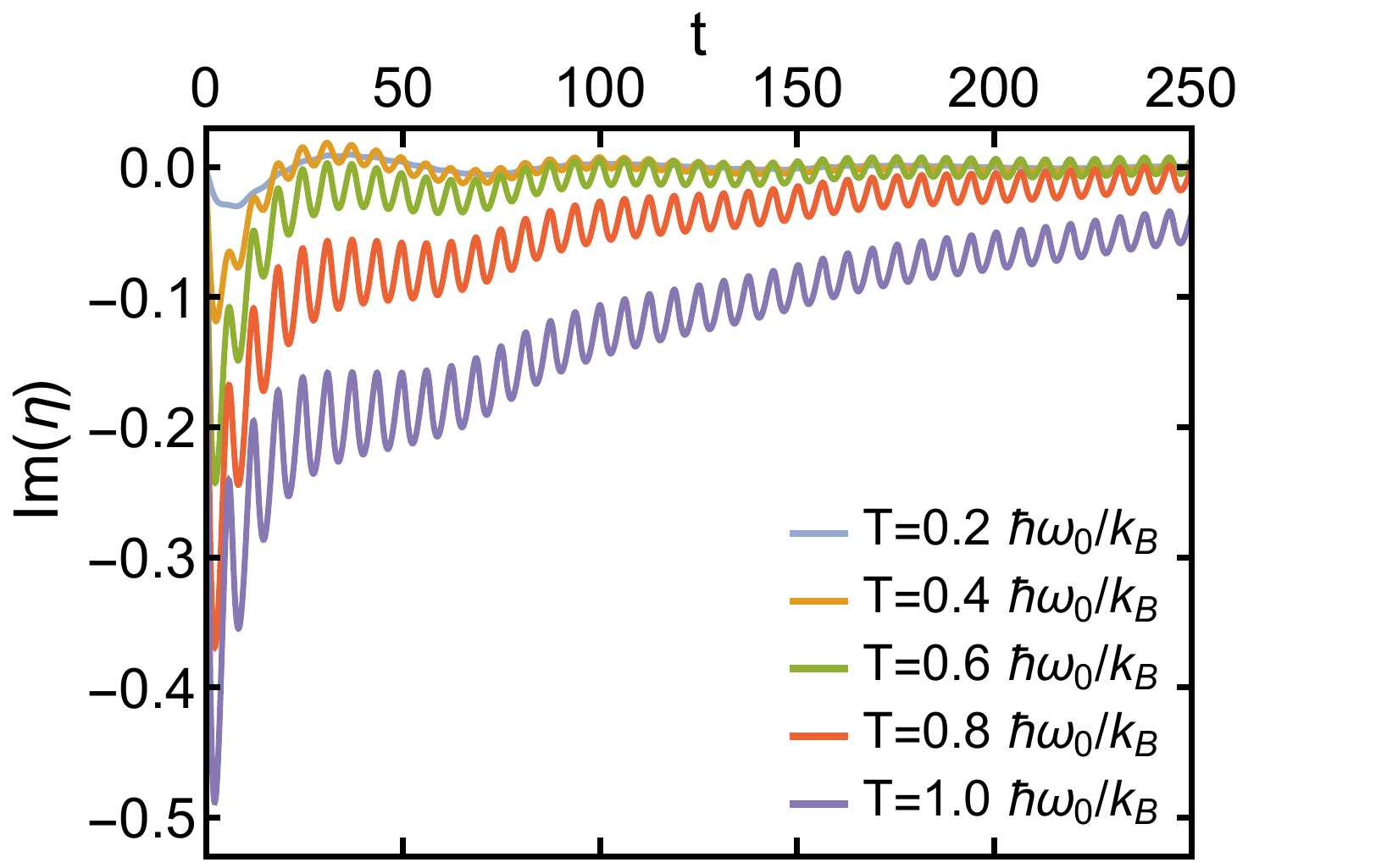}}\\[-2ex]
\subfigure[$s=2$]{\label{fig-PlotR23}
\includegraphics[width=0.25\textwidth]{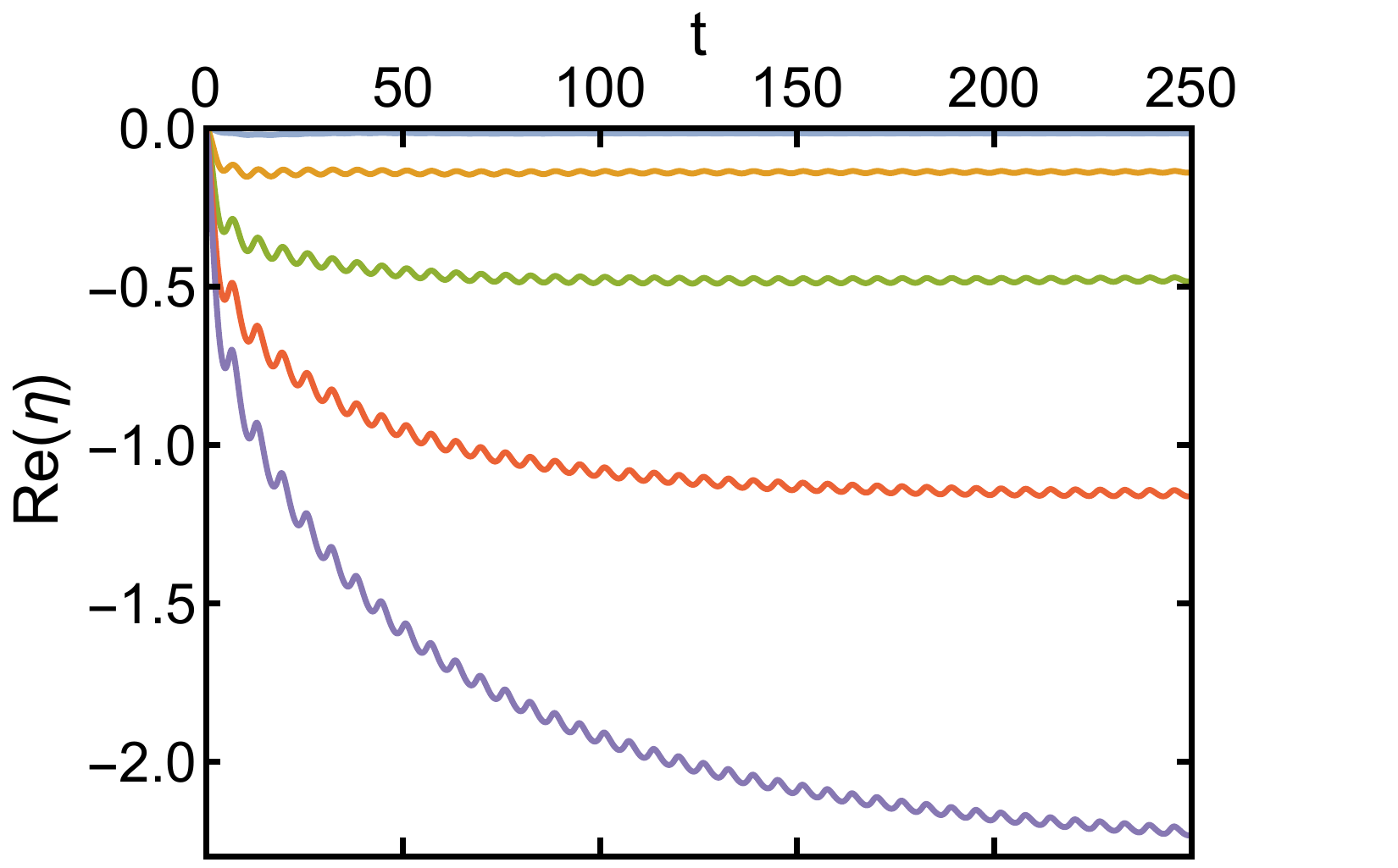}}\hspace{-2.2em}    
\subfigure[$s=2$]{\label{fig-PlotI23}
\includegraphics[width=0.25\textwidth]{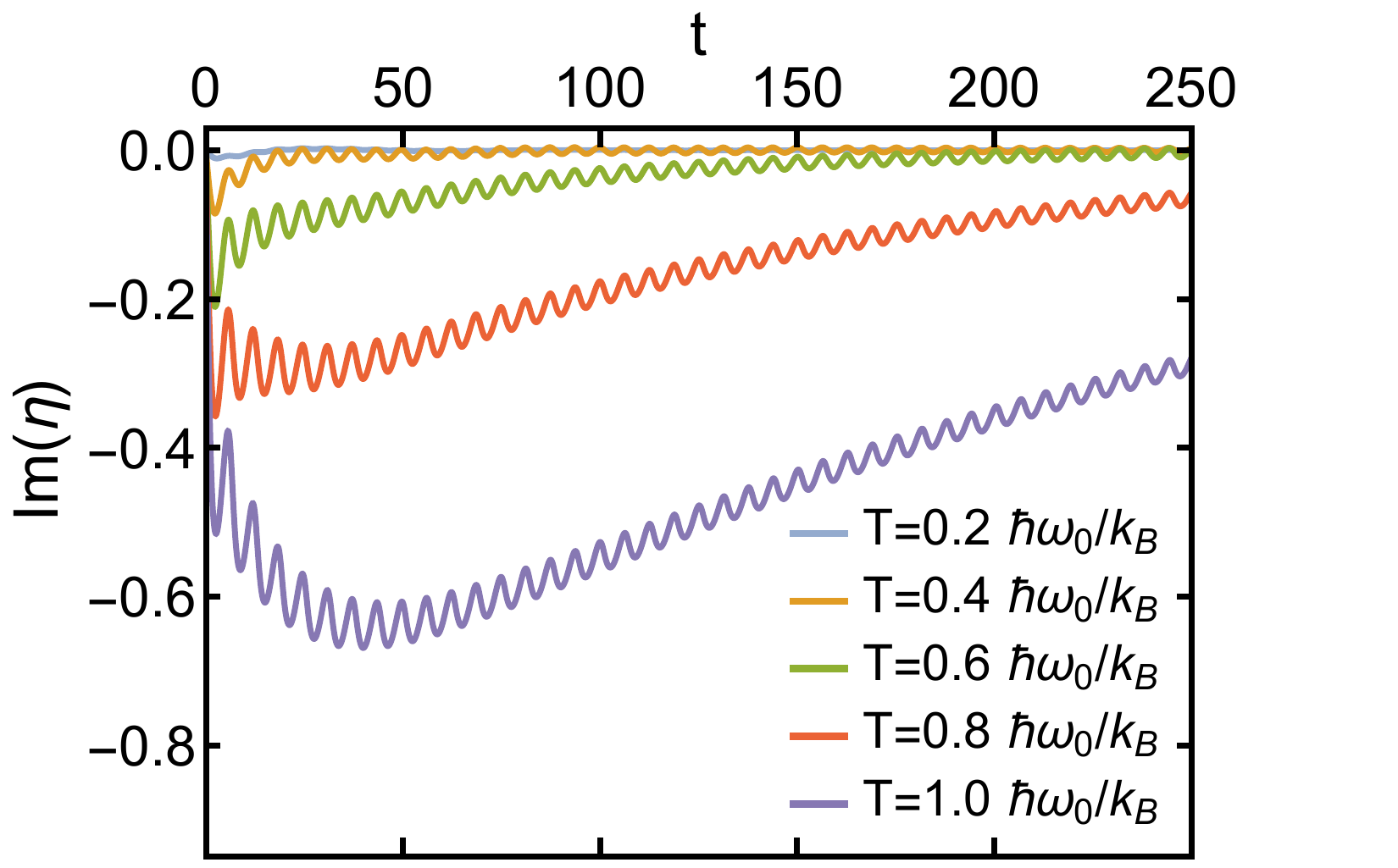}}
\caption
{\small  $\mathrm{Re}(\eta(t))$ and $\mathrm{Im}(\eta(t))$ for $s=0,1,2$ and $T=0.2, 0.4,0.6,0.8,1.0\hbar\omega_{0}/k_{\rm B}$. $\lambda$ and $\omega_{0}$ are both set as 1, and $\sigma$ is set as 0.3.}
\label{fig-zeta}
\end{figure}

 Second, let $\lambda_\omega$ be in the Lorentzian form. That is, 
\begin{align}
	\lambda_\omega =\frac{\lambda}{1+({\omega-\omega_0})^2/{\sigma}^2} \, ,
\end{align}
where $\lambda$ is the maximum coupling strength; $\omega_0$ is the frequency of the mode which is most intensively coupled with the system; and $\sigma$ is the width parameter of the Lorentzian function. In Fig.~\ref{fig-zeta}, we plot the real and imaginary part of $\eta(t)$ with respect to various choices of temperature $T$ and parameter $s$, where have set $\lambda=1$, $\omega_0=1$ and $\sigma=0.3$. Note that for all those choices of parameters, in the long-time asymptotic regime, the real part of $\eta(t)$ approaches to some steady value and the imaginary part of $\eta(t)$ approaches to zero, which indicates that the off-diagonal elements of the system density matrix will \emph{not} decay to zero. 
\blk

At low temperatures ($T=0.2 \hbar \omega_0/k_{\rm B}$, for instance), only low-energy bath modes will be involved in the interaction with the system. 
That is, the dephasing strength is expected to be weak. In fact, as shown in Fig.~\ref{fig-zeta}, both the values of the real and imaginary parts of $\eta(t)$ for $T=0.2 \hbar \omega_0/k_{\rm B}$ are bigger and oscillating less intensively than that of other higher temperatures.

On the other hand, for a given temperature, as the parameter $s$ increases, the contribution from the low-energy bath modes decreases and that of the high-energy bath modes will gradually dominate the dynamics. In terms of the dephasing function $\eta(t)$, the amplitude of its envelope oscillation, which is contributed by the system's interaction with the low-energy bath modes, should decreases with $s$ increases. Indeed, as shown in Fig.~\ref{fig-zeta}, the envelope oscillating for $s=0$ is quite intensive; it becomes less intensive when $s=1$, and almost vanishes when $s=2$.


\blk
Finally, we consider the case that $\lambda_\omega$ is linearly dependent on the frequency of bath modes. That is, $\lambda_{\omega}=f \omega$, where $f$ is a constant. In this case, $\eta(t)$ can be written as \blk
\begin{align}\label{eq-eta-lw}
{\eta} (t) =-\Gamma(1+s)\zeta(2+s) T^{1+s} \bigg(1-\frac{1}{\l({1+{i} f T t}\r)^{1+s}} \bigg)\, ,
\end{align}
where $\zeta(2+s)=\sum_{n=1}^{\infty}{1}/{n^{2+s}}$ denotes the Riemann zeta function. This case will be further explored in Sec.~\ref{sec-dep-inf} for the discussion on the non-Markovian properties of our model.


\blk	
\section{Reproducing the dephasing of a spin}
\label{sec-rep-sb}
	
For a spin-boson pure-dephasing model, its Hamiltonian can be typically written as~\cite{BP02}
\begin{align}\label{eq-spinboson}
H_{\rm tot}^{\rm sb} = \, & \omega_{\rm S}\ketbra{1}{1} + \sum_{k} \omega_k b_k^{\dagger} b_k \nonumber \\ & +\sum_{k} \sigma_z(\tilde{\lambda}_k b_k  + \tilde{\lambda}_k^{*}b_k^{\dagger}) \, ,
\end{align}
where $\sigma_z$ is the Pauli operator and $\lambda_k$ describes the coupling strength. Assume the system state at the initial time is $\rho_{\rm spin}(0)=\begin{pmatrix}
r_{00}     &  r_{01} \\
r_{10}     &  r_{11}
\end{pmatrix}$. It is well-known that the reduced system state has the analytical form~\cite{BP02}: \blk
\begin{align}
\label{bb-sb}
\rho_{\rm spin} (t)=
\begin{pmatrix}
r_{00}      &  r_{01} e^{{i} \omega_{\rm S} t +{\tilde{\eta}}(t)}\\
r_{10} e^{-{i} \omega_{\rm S} t+{\tilde{\eta}}(t)} &  r_{11}
\end{pmatrix} \, ,
\end{align}
where 
\begin{align}
\tilde{\eta}(t)=-4\sum_k { |\tilde{\lambda}_k|^2}\operatorname{coth}\l(\frac{\omega_k}{2T}\r)\frac{1-\operatorname{cos}\omega_k t}{\omega_k^2}\, ,
\end{align}
or (in the continuous limit of the bath modes)
\begin{align}
\tilde{\eta}(t)=-4 \int_{0}^{\infty}{\rm d}\omega g_\omega { |\tilde{\lambda}_\omega|^2}\operatorname{coth}\l(\frac{\omega}{2T}\r)\frac{1-\operatorname{cos}\omega t}{{\omega^2}}\, ,
\end{align}
with $g_\omega$ denoting the density of states of the bath. 

If the initial state of the system in our model~\eqref{eq-hs}-\eqref{eq-Hinter} is constraint to be in the form $\rho_{\rm S}(0)=\begin{pmatrix}
C_{00}     &  C_{01} \\
C_{10}     &  C_{11}
\end{pmatrix}$, where ``0'' and ``1'' represent the particle number of the system, the boson-boson model can reproduce the dephasing of a spin. Specifically, the effective total Hamiltonian in Eqs.~\eqref{eq-hs}-\eqref{eq-Hinter} can be rewritten as
\begin{align}\label{eq-eff}
H_{\rm eff}^{\rm bb}=\omega_{\rm S}\ketbra{1}{1} + \sum _k \omega _k  b_k^{\dagger } b_k+\sum _k \lambda_k \ketbra{1}{1} b_k^{\dagger} b_k \, ,
\end{align}
where $\ket{1}$ denotes the system state with one particle. The system state is then given by 
\begin{align}
\label{bb-spin}
\rho_{\rm S} (t)=
\begin{pmatrix}
C_{00}      &  C_{01} e^{{i} \omega_{\rm S} t +{\eta}^{*}(t)}\\
C_{10} e^{-{i} \omega_{\rm S} t+{\eta}(t)} &  C_{11}
\end{pmatrix} \, .
\end{align}

Quantitatively, Eq.~\eqref{bb-spin} reproduces the result from the spin-boson pure-dephasing model in Eq.~\eqref{bb-sb}, although the mechanisms behind dephasing are different as obviously seen in the Hamiltonians. In fact, for non-zero temperature cases, vacuum fluctuation and thermal fluctuation could coexist in the spin-boson model~\cite{BP02}. While in our boson-boson model, as discussed in Sec.~\ref{sec-2B}, the amplitude of the oscillation vanishes as the temperature approaches zero, meaning that the dephasing is purely induced from the thermal fluctuations in the bath.

\blk
\section{Non-Markovian properties of the boson-boson pure-dephasing model}
\label{sec-non-Mar}

In the previous two sections, we have shown that our model possesses non-Markovian properties. In this section, by employing three non-Markovian criteria, we study the non-Markovianity in our model in more detail.

It is worth emphasizing that there are many different approaches to qualifying non-Markovianity for open quantum systems. In some papers, they are often referred to as ``measures''~\cite{RHP14,BLP16}.  However, while all these approaches are well justified in physics, generally they may fail to reach a consensus when being applied to  a specific  open system~\cite{MJL14}. In fact, there is a quite complicated hierarchical relation  among different approaches~\cite{LHW17}. For this reason, we would refer to them as non-Markovian witnesses in this paper. That is,  they are sufficient conditions for a non-Markovian system, but not necessary. In particular, we consider two well-known criteria based on divisibility~\cite{RHP10} and quantum regression formula~\cite{GSV14}, respectively. Furthermore, we propose our criterion based on the Wigner function~\cite{SZ99}, and study the non-Markovianity in our model with it.
	
\subsection{Divisibility}
\label{sec_Nivisibility}
	
The dynamics of an open quantum system can be described by a family of \emph{dynamical maps} $\{ {\E} (t,0), t \geq 0 \}$. That is,
\begin{align}
\rhos(t) = {\cal E}(t,0) \, \rhos(0) \,  ,
\end{align}
where ${\E}(t,0)$ is defined as
\begin{align}
	{\E}(t,0) X := \tr_{\rm B} \l[  U(t, 0) \l( X \otimes \rho_{\rm B} \r)  U^\dagger (t, 0) \r]
	\end{align}
for any system operator $X$~\cite{LHW17}.  By its definition, a dynamical map is completely positive and trace-preserving (CPTP)~\cite{BP02}. From Eq.~\eqref{eq-densitymatrix}, it is easy to see that the dynamical map for the boson-boson pure-dephasing model is given by
\begin{align}\label{eq-dyn-map}
{\E}(t,0) \rhos(0) = \sum_{j,k=0}^{\infty} C_{jk} e^{-{i}(j-k) \omega_{\rm S} t+\eta((j-k)t)} \ketbra{j}{k}\, ,
\end{align}
where $ \rhos(0) = \sum_{j,k=0}^{\infty} C_{jk} \ketbra{j}{k}$.

A widely employed definition of quantum Markovianity is proposed in Ref.~\cite{RHP10}, where the dynamics of an open quantum system is taken to be Markovian if and only if its dynamical map is \emph{divisible}. That is, for any $t \geq t' \geq 0$, there exists a CPTP map ${\cal Q}(t,t')$ such that
\begin{align}
{\cal E}(t,0) = {\cal Q}(t,t') {\cal E}(t',0) \,  .
\end{align}
In Ref.~\footnote[1]{Manuscript in preparation.}, 
we show that for our model, a sufficient and necessary condition for ${\E}(t,0)$ being divisible is that the dynamical map of an arbitrary subsystem is divisible. To be specific, consider a subsystem described by a density matrix $\rho_n (0) =\sum_{j,k=0}^{n-1} \rho_{jk} \ketbra{j}{k}$, where $\ket{j}$ and $\ket{k}$ denote the number states and $n \geq 2$ is an integer that is not grater than the dimension of the system.  The dynamics of that subsystem can then be described by a family of CPTP maps~$\{ {\cal E}_n(t, 0), t \geq 0 \}$ where $\E_n(t,0)$ satisfies that
\begin{align}\label{eq-dyn-map-d}
\E_n(t,0) \sum_{j,k=0}^{n-1} \rho_{jk} \ketbra{j}{k} = \sum_{j,k=0}^{n-1} \rho_{  jk} e^{-{i}(  j - k ) \omega_{\rm S} t+\eta((  j - k )t)} \ketbra{  j}{  k}\, .
\end{align}
The corresponding master equation is
\begin{equation}
\label{eq-ME-div1}
\dot{\rho}_n (t) =-\mathrm{i}[H_{\rm S}, \rho_n (t) ]+\sum_{j,k=0}^{n-1}\dot{\eta}((j-k)t)\ketbra{j}{j} \rho_n \ketbra{k}{k}\, .
\end{equation}
Following the method developed in Ref.~\cite{MJL14}, we can rewrite Eq.~\eqref{eq-ME-div1} as
\begin{align}
\dot{\rho}_n=-\mathrm{i}[H_{\rm S}', \rho_n]+\sum_{p,q=1}^{n^2-1} d_{pq}(t)\Big(G_p \rho_n G_q-\frac{1}{2}\l\{G_q G_p, \rho_n\r\}\Big)\, ,
\end{align}
where $H_{\rm S}'=H_{\rm S}'^{\dagger}$; $\{G_p: p=0,1,\cdots, n^2-1\}$ satisfies
\begin{equation}
G_0=\frac{\mathbb{I}_n}{\sqrt{n}}, \: \: \: G_p=G_p^{\dagger},\: \: \: \operatorname{Tr}\l[G_p G_q\r]=\delta_{pq}\, ;
\end{equation}
and 
\begin{align}
d_{pq}(t)=\sum_{j,k=0}^{n-1} \dot{\eta}((j-k)t) \bra{j}G_p\ket{j} \bra{k}G_q\ket{k}\, .
\end{align}
Note that $\{d_{pq}(t):p,q\in\{1,2,\cdots,n^2-1\}$ form the $(n^2-1)\times(n^2-1)$ Hermitian matrix $\bm{d}_n(t)$ known as the \emph{decoherence matrix}~\cite{MJL14}. ${\E}_n(t,0)$ is divisible if and only if $\bm{d}_n(t)$ is positive-semidefinite, i.e., $\bm{d}_n(t)\geq0$~\cite{MJL14}. 

By choosing a specific representation, the $G_p$'s can be classified as
\begin{align}
G_l^{\rm d}&=\frac{\operatorname{diag}\{1,\cdots,1,-l,0,\cdots,0\}}{\sqrt{l(l+1)}}  \;\;  (1\leq l\leq n-1)\, ;\\
G_{jk}^{\rm s}&=\frac{1}{\sqrt{2}}(\ketbra{j}{k}+\ketbra{k}{j})   \qquad\quad (0\leq k < j\leq n-1)\, ; \\
G_{jk}^{\rm a}&=\frac{i}{\sqrt{2}}(\ketbra{j}{k}-\ketbra{k}{j})  \qquad\quad (0\leq k < j\leq n-1)\, ;
\end{align}
where the superscripts  stand for diagonal, symmetric, and anti-symmetric respectively. 
Sort $G_l^{\rm d}$ in the ascending order of $l$, and $G_{mn}^{\rm s}$ ($G_{mn}^{\rm a}$) in the ascending order of $m$ and $n$ respectively. Then, in the decoherence matrix $\bm{d}_n(t)$, only the upper-left $(n-1) \times (n-1)$ block $\bm{d}_n^{\rm B}(t)$ is non-trivial, i.e., all the other matrix elements are 0. Consequently, $\bm{d}_n(t)\geq 0$ is equivalent to $\bm{d}_n^{\rm B}(t)\geq 0$. 

As a simple example, consider the case $n=2$. $\bm{d}_n^{\rm B}(t)$ is thus simply a scalar as \blk
\begin{align}
\bm{d}^{\rm B}_2(t)=\l[-\operatorname{Re}(\dot{\eta}(t))\r]_{1\times 1}\, .
\end{align}
Therefore,  ${\cal E}_2(t,0)$ is divisible if and only if ${\rm Re}(\dot{\eta}(t)) \leq 0$.

Another example is when $n=3$,  one has
\begin{align}
\bm{d}_3^{\rm B}(t)=
\begin{pmatrix}{-\operatorname{Re}(\dot{\eta}(t))}&{\frac{-\operatorname{Re}(\dot{\eta}(t))+2\dot{\eta}^{*}(t)-\dot{\eta}^{*}(2t)}{\sqrt{3}}} \\
{\frac{-\operatorname{Re}(\dot{\eta}(t))+2\dot{\eta}(t)-\dot{\eta}(2t)}{\sqrt{3}}} &{-\frac{\operatorname{Re}(\dot{\eta}(t))+2\operatorname{Re}(\dot{\eta}(2t))}{\sqrt{3}}} 
\end{pmatrix}\, .
\end{align}
It is positive-semidefinite if and only if $\operatorname{Re}(\dot{\eta}(t))\leq 0$ and $\l|\bm{d}^{\rm B}_3(t)\r|\geq 0$~\cite{A97}. Note that $\l|\bm{d}_3^{\rm B}(t)\r|$ is related to the imaginary part of $\eta(t)$. In this sense, divisibility is also related to the phase information of the system.

For $n>3$, one can also derive the corresponding divisibility condition. All these conditions form a hierarchy~\footnotemark[1]. If for some $n$, the condition is violated, the dynamics characterized by $\l\{\E(t,0),t\geq0 \r\}$ is non-Markovian. Generally speaking, the real part of $\eta(t)$ is a non-monotonic function, thus violating the divisibility condition given by the case $n=2$. Therefore, our model possesses non-Markovianity. 


\subsection{Correlation function and  quantum regression formula}
\label{sec_correlation function}
	
Calculations of correlation functions of open quantum systems are usually rather complex as the total Hilbert spaces are quite involved~\cite{BP02}. Under certain conditions, it is possible to evaluate those correlation functions only on the system Hilbert space, which therefore greatly reduces the computation complexity. This is well-known as the {quantum regression formula}~\cite{HC991,GZ04}. Furthermore, it has been shown that the violation of {quantum regression formula}, to some extent, implies the system dynamics is non-Markovian~\cite{GSV14,ALT15}. That is, for any two system operators $A$ and $B$, the system dynamics is Markovian if and only if the correlation function $\langle A(t) B(t+\tau) \rangle$ satisfies that for $\tau\geq 0$,
\begin{align}
	\label{eq_QRF}
	\langle A(t) B(t+\tau) \rangle = \tr_{\rm S} \l[ B {\E}(t+\tau,t) \l[ \rhos(t) A \r] \r]\, ,
	\end{align}
where ${\E}(t+\tau,t)$ is the dynamical map within the time interval from $t$ to $t+\tau$~\cite{GZ04}.
	
In this subsection, we will use this criterion to demonstrate the non-Markovianity of the boson-boson pure-dephasing model.  We will first calculate the two-time correlation function $\langle b (t) b (t+ \tau)\rangle$, and then compare it with the result from quantum regression formula Eq.~\eqref{eq_QRF}.

The time evolution of the system annihilation operator $b$ can be easily evaluated in the Heisenberg picture. It follows from Eqs.~\eqref{eq-hs}-\eqref{eq-Hinter} that
	\begin{align}
	b (t)  =e^{-{i}({\omega}_{\rm S}+\sum_k {\lambda_k {\hat{n}}_k})t}b \, .
	\end{align}
The correlation function $\langle b (t) b (t+ \tau)\rangle$ then can be directly calculated as
	\begin{align}
	\label{eq_exact_correlation_fucntion}
	\langle b (t) b (t+ \tau)\rangle &= \tr \l[ b (t) b (t+ \tau) \rho_{\rm S}(0)\otimes\rho_B \r] \nonumber \\
	&= e^{-{i} \omega_{\rm S} (2t+\tau)} \langle e^{-{i} (2t+\tau) \sum_k{\lambda_k \hat{n}_k}} \rangle \langle b^2 \rangle  \nonumber  \\
	&=e^{-{i} \omega_{\rm S}  (2t+\tau)} e^{{\eta}(2t+\tau)}  \langle b^2 \rangle \, ,
	\end{align}
where $\langle b^2 \rangle=\tr_{\rm S}\{\rho_{\rm S}(0)b^2\}$.
	%
	%
	%
	
Now we proceed with the calculation of {quantum regression formula}. First note that for $t\geq 0$ and $\tau \geq 0$, we have
\begin{align}\label{eq-dyn-map-X}
{\E}(t+\tau,t) X  = \sum_{m,n}  X_{mn} e^{- {i} (m - n) \omega_{\rm S}\tau } e^{ \eta\l( (m-n)\tau \r)  } \, ,
\end{align}
where  $X = \sum_{m,n} X_{mn} |m\rangle\langle n |$. According to the quantum regression formula in Eq.~\eqref{eq_QRF}, we have
	\begin{align}
	\label{de_QRF1}
	\langle b (t) b (t+ \tau) \rangle_{\rm QRF} & = \tr_{\rm S}  \l\{  b  {\E}(t+\tau,t) \l[ \rhos(t) b \r] \r\} \,  .
	\end{align}
Because $\rhos(t)$ is in the form of Eq.~\eqref{eq-densitymatrix}, and ${\E}(t+\tau,t)$ satisfies Eq.~\eqref{eq-dyn-map-X}, Eq.~\eqref{de_QRF1} can be transformed to
	\begin{align}
	& \langle b (t) b (t+ \tau) \rangle_{\rm QRF}   = e^{-{i} \omega_{\rm S}  (2t+\tau)} e^{{\eta}(2t)+\eta(\tau)}  \langle b^2 \rangle \,.
	\end{align}

Obviously, $\langle b (t) b (t+ \tau) \rangle = \langle b (t) b (t+ \tau) \rangle_{\rm QRF}$ if and only if
\begin{align}
{\eta}(2t+\tau)={\eta}(2t)+\eta(\tau)\, ,
\end{align}
i.e., $\eta(t)$ is linear in $t$ and $\eta(0)=0$.
Generally speaking, $\eta(t)$ in our model does not satisfy this condition, meaning that our model is intrinsically a non-Markovian system in the sense of violating the quantum regression formula~\cite{LHW17}.

\subsection{Non-Markovianity characterized in terms of Wigner function}
\label{sec-dep-inf}
Quantum systems with continuous variables are often described by the Wigner function~\cite{BV05}. In this subsection, we characterize the memory effect in our boson-boson model using  the Wigner function. \blk

The Wigner function of a density matrix $\rho(t)$ is usually defined by~\cite{WM07}
	\begin{align}\label{eq-wig-def}
	W(\gamma, t)=\frac{1}{{\pi}^2} \int \mathrm{d^{2}} \xi e^{{\xi}^* \gamma -\xi {\gamma}^*} \mathrm{Tr}\{ \rho(t) e^{\xi b^{\dagger}-{\xi}^* b}\}\,.
	\end{align}
In Ref.~\cite{KZ04}, the notion called {quantumness} with respect to the Wigner function $W(\gamma)$ is defined as
\begin{align}\label{eq-quan}
\delta(t) = \frac{1}{2} \left(\int\mathrm{d^{2}} \gamma \left|{W}(\gamma, t)\right| -1\right) \, ,
\end{align}
which can be interpreted as the summation of all the negative probabilities. 
\blk
It has been shown in Ref.~\cite{SPID05} that the decreasing of $\delta(t)$ is a signature of decoherence. 
In the following discussion, let us reconsider the last scenario discussed in Sec.~\ref{sec-2B} where $g_\omega=\omega^s$ and $\lambda_{\omega}=f \omega$. We show that the changes in quantumness do capture the memory effect in our model and the result is consistent with the discussion on the dephasing function $\eta(t)$ in Sec.~\ref{sec-2B}.

\begin{figure}
	\centering
	\subfigure[]{\label{2a}
		\includegraphics[width=0.22\textwidth]{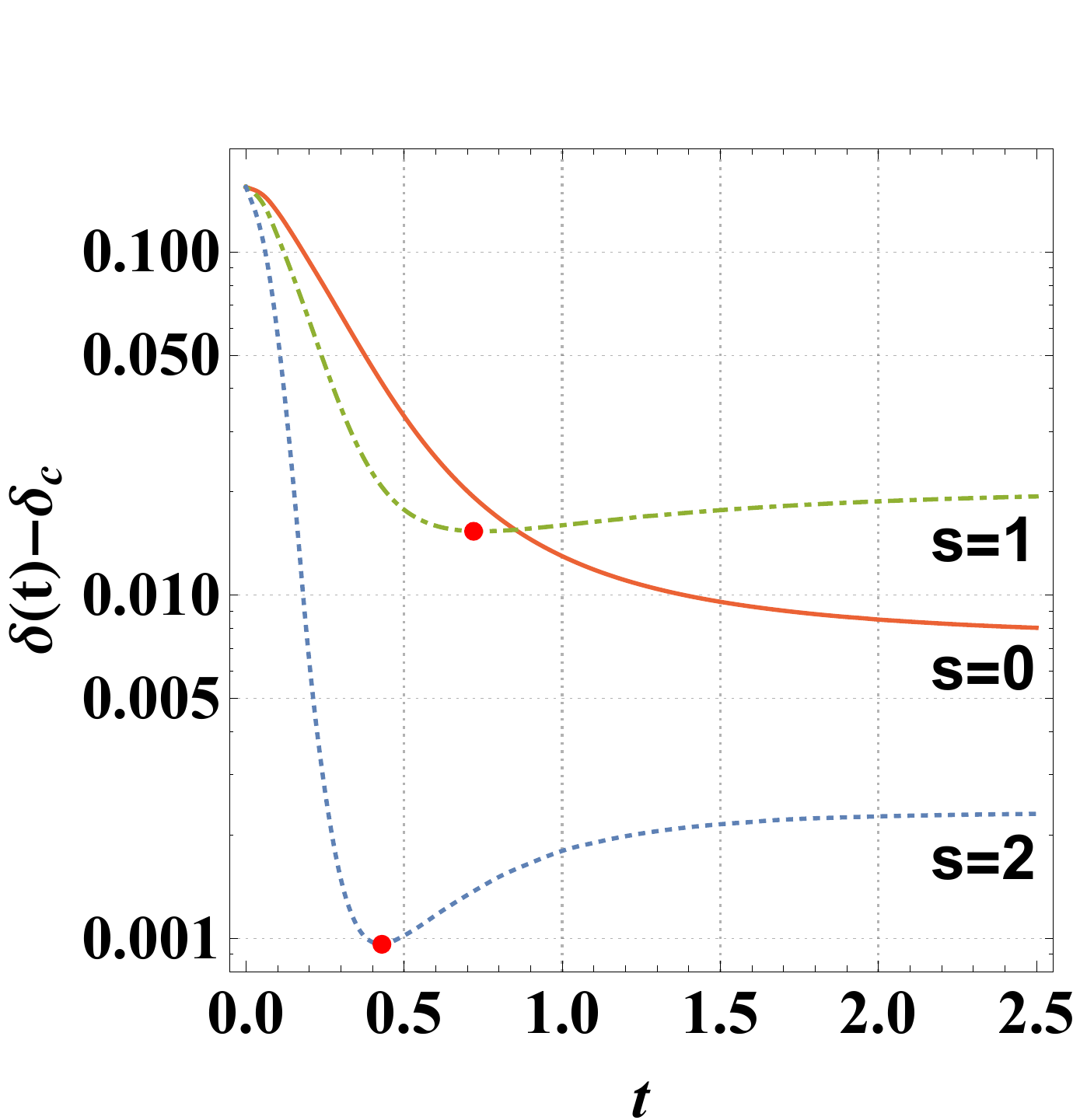}}\hspace{-0.4em}    
	\subfigure[]{\label{2b}
		\includegraphics[width=0.22\textwidth]{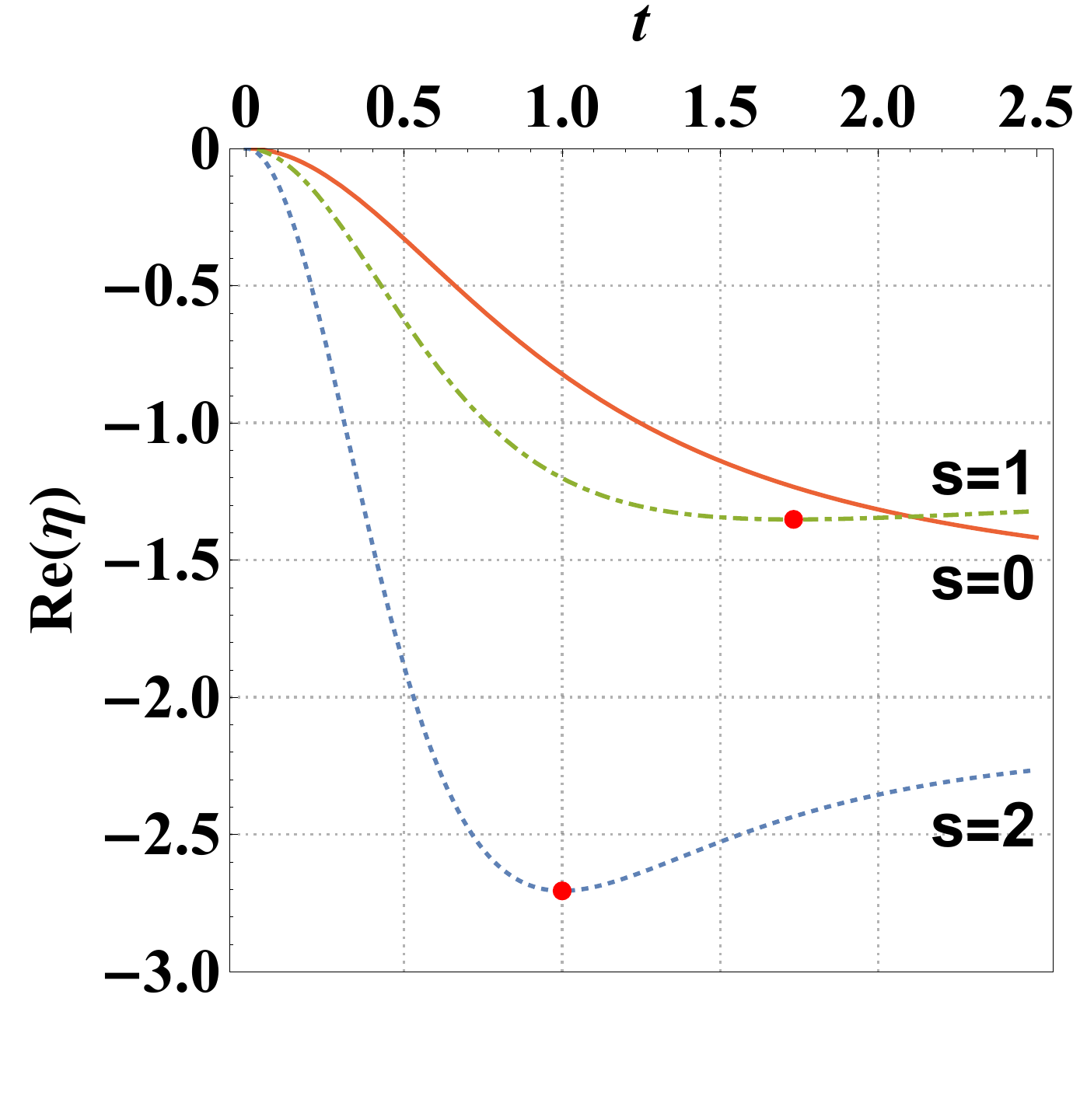}}\\[-2ex]
	\caption
	{\small Plots of $\delta(t)-\delta_c$ and $\operatorname{Re}(\eta(t))$. The initial state is chosen to be the Schr\"{o}dinger cat state $(|\alpha\rangle+\left|-\alpha \right\rangle)/\mathcal{N}$ with $\alpha=1.5$. $f=1$ and $T=1$.  The red dot in the plots represent the minimum of the corresponding function.}
	\label{fig-noncla-evo}
\end{figure}

In particular, we choose \blk the initial state as the Schr\"{o}dinger cat state $(|\alpha\rangle+\left|-\alpha \right\rangle)/\mathcal{N}$, where $|\alpha\rangle$ and $\left|-\alpha\right\rangle$ denote coherent states, $\alpha\geq0$, and $\mathcal{N}=\sqrt{2(1+e^{-2 {\alpha}^2})}$~\cite{WM07}. \blk  The corresponding Wigner function then reads
\begin{widetext}
	\begin{equation}\label{eq-quan-cat}
	W(\gamma,t)=\frac{2 e^{-2|\gamma|^2}}{\pi(1+e^{-2 {\alpha}^2})}\bigg[e^{-2 {\alpha}^2} \operatorname{I_0} (x)+ \operatorname{J_0}(x)
	+\sum_{k=1}^{\infty}{\Big(e^{-2 {\alpha}^2} \operatorname{I}_{2k}(x)+ \operatorname{J}_{2k}(x)\Big)\Big(e^{{\eta}(2kt)-i2k(\omega_0 t+\arg \gamma)}+\operatorname{c.c.}\Big)}\bigg] \, , 
	\end{equation}
\end{widetext}
where $x=4\alpha|\gamma|$; $\operatorname{I}_0 (x)$ and $\operatorname{I}_{2k} (x)$ denote the modified Bessel functions of the first kind~\cite{O10}; $\operatorname{J}_0(x)$ and $\operatorname{J}_{2k} (x)$ denote the Bessel functions of the first kind~\cite{O10}; and $\eta(t)$ satisfies Eq.~\eqref{eq-eta-lw}.  We plot the corresponding quantumness in Fig.~\ref{2a}, as a function of time $t$ (see Eq.~\eqref{eq-quan}). Note that for better illustrating the properties of quantumness, we have subtracted a constant $\delta_{\rm c}$ from it, which denotes the contribution from time-independent terms in the Wigner function, i.e., 
\begin{align}
\delta_{\rm c} = \frac{1}{2} \left(\int {\rm d}^{2} \gamma \left|{W}_{\rm c}(\gamma)\right| -1\right) \, ,
\end{align}
where $W_{\rm c}$ is the first two terms in Eq.~\eqref{eq-quan-cat}.

In the numerical simulation, we have set $\lambda_\omega = \omega$, $T$ to be $1$, and vary the parameter $s$, as shown in Fig.~\ref{2a}. $\delta(t)$ is a non-monotonic function when $s$ is set to 1 or 2, indicating there should be information back flow between the system and bath which leads to that partial recurrence in quantumness. Such non-monotonic property therefore can be taken as a signature of quantum non-Markovianity. When $s = 0$, $\delta(t)$ decays monotonically. However, it is not sufficient to justify that the dynamics is Markovian as we have merely verified it with one initial state~\cite{RHP14,LHW17}.

The conclusion we draw here is further supported by revisiting the discussion on the divisibility of our model in Sec.~\ref{sec_Nivisibility}. Recall that we have shown the dynamics in our model is non-Markovian (in the sense of divisibility) if the real part of $\eta(t)$ is a non-monotonic function. In fact, with exact the same setting of parameters, we find that the properties of ${\rm Re}(\eta)$ are consistent with $\delta(t)$ as shown in Fig.~\ref{2b}. In particular, they both show non-monotonic property when $s$ is set to 1 and 2, and decay monotonically in the case of $s = 0$.

\section{Conclusions}
\label{sec-con}

In this paper, we propose a boson-boson pure-dephasing model where the system and the bath interact through cross-Kerr interaction. It is shown that the system  dynamics can be characterized by  the dephasing function $\eta(t)$, which is induced by the thermal fluctuations in the bath. \blk By studying the properties of the function for several typical cases, we illustrate that there exist memory effects in the dynamics. Moreover, our model can reproduce the pure-dephasing dynamics of a spin. We show that the dynamics differs from the pure-dephasing dynamics of the well-known spin-boson model by the dephasing mechanisms, therefore showing different dependence on the temperature $T$, the density of states $g_\omega$ and the coupling strength function $\lambda_\omega$. \blk 

We also study the non-Markovianity with respect to two different definitions, i.e., divisibility and quantum regression formula. The condition of divisibility is generally characterized by a set of inequalities. Specifically, for a qubit system, this condition is equivalent to the monotonic decay of the moduli of the off-diagonal elements of the system density matrix.  We also show that our model does not satisfy the so-called quantum regression formula. As this is stronger concept of quantum Markovianity than divisibility, we could regard that our model is generically non-Markovian. We also explore the possibility of characterizing quantum non-Markovianity from the perspective of quantumness, a concept derived from Wigner function. For our model, we demonstrate this approach shows a consistent result in witnessing non-Markovianity comparing to divisibility.  However, whether the result holds for general cases could be the work of future study.

Finally, it is worth pointing out that  the cross-Kerr interaction can be interpreted as the collision interaction that does not exchange energy or momentum. This means our model is a simplification of more general bosonic collisional systems. In specific, the fundamental mechanism of Bose-Einstein condensation is colliding interaction~\cite{M13}. If the condensed part is considered as a system and non-condensed part as a bath, the interaction  Hamiltonian is similar to our model. In fact, for future study, by allowing momenta-exchange terms in our model, it may provide a new perspective in characterizing the Bose-Einstein condensation.

\acknowledgments
We gratefully thank Wei-Min Zhang  for his insight in modeling dephasing process with  boson-boson systems and for offering us helpful advices in this work. We also thank Pei-Yun Yang, Ping-Yuan Lo, Hon-Lam Lai, I-Chi Chen and Yu-Wei Huang for helpful discussions. F.L.X. and Z.B.C. were supported by the National Natural Science Foundation of China (Grant No. 61125502) and the CAS. L.L. was supported by the ARC Centre of Excellence CE110001027.
	
	
	
	\bibliographystyle{apsrev4-1}
	\bibliography{Pure-dephasing}

\begin{thebibliography}{49}%
\makeatletter
\providecommand \@ifxundefined [1]{%
 \@ifx{#1\undefined}
}%
\providecommand \@ifnum [1]{%
 \ifnum #1\expandafter \@firstoftwo
 \else \expandafter \@secondoftwo
 \fi
}%
\providecommand \@ifx [1]{%
 \ifx #1\expandafter \@firstoftwo
 \else \expandafter \@secondoftwo
 \fi
}%
\providecommand \natexlab [1]{#1}%
\providecommand \enquote  [1]{``#1''}%
\providecommand \bibnamefont  [1]{#1}%
\providecommand \bibfnamefont [1]{#1}%
\providecommand \citenamefont [1]{#1}%
\providecommand \href@noop [0]{\@secondoftwo}%
\providecommand \href [0]{\begingroup \@sanitize@url \@href}%
\providecommand \@href[1]{\@@startlink{#1}\@@href}%
\providecommand \@@href[1]{\endgroup#1\@@endlink}%
\providecommand \@sanitize@url [0]{\catcode `\\12\catcode `\$12\catcode
  `\&12\catcode `\#12\catcode `\^12\catcode `\_12\catcode `\%12\relax}%
\providecommand \@@startlink[1]{}%
\providecommand \@@endlink[0]{}%
\providecommand \url  [0]{\begingroup\@sanitize@url \@url }%
\providecommand \@url [1]{\endgroup\@href {#1}{\urlprefix }}%
\providecommand \urlprefix  [0]{URL }%
\providecommand \Eprint [0]{\href }%
\providecommand \doibase [0]{http://dx.doi.org/}%
\providecommand \selectlanguage [0]{\@gobble}%
\providecommand \bibinfo  [0]{\@secondoftwo}%
\providecommand \bibfield  [0]{\@secondoftwo}%
\providecommand \translation [1]{[#1]}%
\providecommand \BibitemOpen [0]{}%
\providecommand \bibitemStop [0]{}%
\providecommand \bibitemNoStop [0]{.\EOS\space}%
\providecommand \EOS [0]{\spacefactor3000\relax}%
\providecommand \BibitemShut  [1]{\csname bibitem#1\endcsname}%
\let\auto@bib@innerbib\@empty
\bibitem [{\citenamefont {Arndt}\ \emph {et~al.}(1999)\citenamefont {Arndt},
  \citenamefont {Nairz}, \citenamefont {Vos-Andreae}, \citenamefont {Keller},
  \citenamefont {Van~der Zouw},\ and\ \citenamefont {Zeilinger}}]{ANV09}%
  \BibitemOpen
  \bibfield  {author} {\bibinfo {author} {\bibfnamefont {M.}~\bibnamefont
  {Arndt}}, \bibinfo {author} {\bibfnamefont {O.}~\bibnamefont {Nairz}},
  \bibinfo {author} {\bibfnamefont {J.}~\bibnamefont {Vos-Andreae}}, \bibinfo
  {author} {\bibfnamefont {C.}~\bibnamefont {Keller}}, \bibinfo {author}
  {\bibfnamefont {G.}~\bibnamefont {Van~der Zouw}}, \ and\ \bibinfo {author}
  {\bibfnamefont {A.}~\bibnamefont {Zeilinger}},\ }\href@noop {} {\bibfield
  {journal} {\bibinfo  {journal} {Nature (London)}\ }\textbf {\bibinfo {volume}
  {401}},\ \bibinfo {pages} {680} (\bibinfo {year} {1999})}\BibitemShut
  {NoStop}%
\bibitem [{\citenamefont {Horodecki}\ \emph {et~al.}(2009)\citenamefont
  {Horodecki}, \citenamefont {Horodecki}, \citenamefont {Horodecki},\ and\
  \citenamefont {Horodecki}}]{HHH09}%
  \BibitemOpen
  \bibfield  {author} {\bibinfo {author} {\bibfnamefont {R.}~\bibnamefont
  {Horodecki}}, \bibinfo {author} {\bibfnamefont {P.}~\bibnamefont
  {Horodecki}}, \bibinfo {author} {\bibfnamefont {M.}~\bibnamefont
  {Horodecki}}, \ and\ \bibinfo {author} {\bibfnamefont {K.}~\bibnamefont
  {Horodecki}},\ }\href@noop {} {\bibfield  {journal} {\bibinfo  {journal}
  {Rev. Mod. Phys.}\ }\textbf {\bibinfo {volume} {81}},\ \bibinfo {pages} {865}
  (\bibinfo {year} {2009})}\BibitemShut {NoStop}%
\bibitem [{\citenamefont {Zurek}(2003)}]{Z03}%
  \BibitemOpen
  \bibfield  {author} {\bibinfo {author} {\bibfnamefont {W.~H.}\ \bibnamefont
  {Zurek}},\ }\href@noop {} {\bibfield  {journal} {\bibinfo  {journal} {Rev.
  Mod. Phys.}\ }\textbf {\bibinfo {volume} {75}},\ \bibinfo {pages} {715}
  (\bibinfo {year} {2003})}\BibitemShut {NoStop}%
\bibitem [{\citenamefont {Gardiner}\ and\ \citenamefont {Zoller}(2004)}]{GZ04}%
  \BibitemOpen
  \bibfield  {author} {\bibinfo {author} {\bibfnamefont {C.}~\bibnamefont
  {Gardiner}}\ and\ \bibinfo {author} {\bibfnamefont {P.}~\bibnamefont
  {Zoller}},\ }\href@noop {} {\emph {\bibinfo {title} {{Quantum Noise: A
  Handbook of Markovian and Non-Markovian Quantum Stochastic Methods with
  Applications to Quantum Optics}}}},\ \bibinfo {edition} {4th}\ ed.\ (\bibinfo
   {publisher} {Springer},\ \bibinfo {address} {Berlin},\ \bibinfo {year}
  {2004})\ p.\ \bibinfo {pages} {450}\BibitemShut {NoStop}%
\bibitem [{\citenamefont {Auffeves}\ \emph {et~al.}(2009)\citenamefont
  {Auffeves}, \citenamefont {G{\'e}rard},\ and\ \citenamefont
  {Poizat}}]{AGP09}%
  \BibitemOpen
  \bibfield  {author} {\bibinfo {author} {\bibfnamefont {A.}~\bibnamefont
  {Auffeves}}, \bibinfo {author} {\bibfnamefont {J.-M.}\ \bibnamefont
  {G{\'e}rard}}, \ and\ \bibinfo {author} {\bibfnamefont {J.-P.}\ \bibnamefont
  {Poizat}},\ }\href@noop {} {\bibfield  {journal} {\bibinfo  {journal} {Phys.
  Rev. A}\ }\textbf {\bibinfo {volume} {79}},\ \bibinfo {pages} {053838}
  (\bibinfo {year} {2009})}\BibitemShut {NoStop}%
\bibitem [{\citenamefont {Naesby}\ \emph {et~al.}(2008)\citenamefont {Naesby},
  \citenamefont {Suhr}, \citenamefont {Kristensen},\ and\ \citenamefont
  {M{\o}rk}}]{NSK08}%
  \BibitemOpen
  \bibfield  {author} {\bibinfo {author} {\bibfnamefont {A.}~\bibnamefont
  {Naesby}}, \bibinfo {author} {\bibfnamefont {T.}~\bibnamefont {Suhr}},
  \bibinfo {author} {\bibfnamefont {P.~T.}\ \bibnamefont {Kristensen}}, \ and\
  \bibinfo {author} {\bibfnamefont {J.}~\bibnamefont {M{\o}rk}},\ }\href@noop
  {} {\bibfield  {journal} {\bibinfo  {journal} {Phys. Rev. A}\ }\textbf
  {\bibinfo {volume} {78}},\ \bibinfo {pages} {045802} (\bibinfo {year}
  {2008})}\BibitemShut {NoStop}%
\bibitem [{\citenamefont {Pfanner}\ \emph {et~al.}(2008)\citenamefont
  {Pfanner}, \citenamefont {Seliger},\ and\ \citenamefont
  {Hohenester}}]{PSH08}%
  \BibitemOpen
  \bibfield  {author} {\bibinfo {author} {\bibfnamefont {G.}~\bibnamefont
  {Pfanner}}, \bibinfo {author} {\bibfnamefont {M.}~\bibnamefont {Seliger}}, \
  and\ \bibinfo {author} {\bibfnamefont {U.}~\bibnamefont {Hohenester}},\
  }\href@noop {} {\bibfield  {journal} {\bibinfo  {journal} {Phys. Rev. B}\
  }\textbf {\bibinfo {volume} {78}},\ \bibinfo {pages} {195410} (\bibinfo
  {year} {2008})}\BibitemShut {NoStop}%
\bibitem [{\citenamefont {Unruh}(1995)}]{U95}%
  \BibitemOpen
  \bibfield  {author} {\bibinfo {author} {\bibfnamefont {W.~G.}\ \bibnamefont
  {Unruh}},\ }\href@noop {} {\bibfield  {journal} {\bibinfo  {journal} {Phys.
  Rev. A}\ }\textbf {\bibinfo {volume} {51}},\ \bibinfo {pages} {992} (\bibinfo
  {year} {1995})}\BibitemShut {NoStop}%
\bibitem [{\citenamefont {Palma}\ \emph {et~al.}(1996)\citenamefont {Palma},
  \citenamefont {Suominen},\ and\ \citenamefont {Ekert}}]{PSE96}%
  \BibitemOpen
  \bibfield  {author} {\bibinfo {author} {\bibfnamefont {G.~M.}\ \bibnamefont
  {Palma}}, \bibinfo {author} {\bibfnamefont {K.-A.}\ \bibnamefont {Suominen}},
  \ and\ \bibinfo {author} {\bibfnamefont {A.~K.}\ \bibnamefont {Ekert}},\ }in\
  \href@noop {} {\emph {\bibinfo {booktitle} {Proceedings of the Royal Society
  of London A: Mathematical, Physical and Engineering Sciences}}},\ Vol.\
  \bibinfo {volume} {452}\ (\bibinfo {organization} {The Royal Society},\
  \bibinfo {year} {1996})\ pp.\ \bibinfo {pages} {567--584}\BibitemShut
  {NoStop}%
\bibitem [{\citenamefont {Duan}\ and\ \citenamefont {Guo}(1998)}]{DG98}%
  \BibitemOpen
  \bibfield  {author} {\bibinfo {author} {\bibfnamefont {L.-M.}\ \bibnamefont
  {Duan}}\ and\ \bibinfo {author} {\bibfnamefont {G.-C.}\ \bibnamefont {Guo}},\
  }\href@noop {} {\bibfield  {journal} {\bibinfo  {journal} {Phys. Rev. A}\
  }\textbf {\bibinfo {volume} {57}},\ \bibinfo {pages} {737} (\bibinfo {year}
  {1998})}\BibitemShut {NoStop}%
\bibitem [{\citenamefont {Reina}\ \emph {et~al.}(2002)\citenamefont {Reina},
  \citenamefont {Quiroga},\ and\ \citenamefont {Johnson}}]{RQJ02}%
  \BibitemOpen
  \bibfield  {author} {\bibinfo {author} {\bibfnamefont {J.~H.}\ \bibnamefont
  {Reina}}, \bibinfo {author} {\bibfnamefont {L.}~\bibnamefont {Quiroga}}, \
  and\ \bibinfo {author} {\bibfnamefont {N.~F.}\ \bibnamefont {Johnson}},\
  }\href@noop {} {\bibfield  {journal} {\bibinfo  {journal} {Phys. Rev. A}\
  }\textbf {\bibinfo {volume} {65}},\ \bibinfo {pages} {032326} (\bibinfo
  {year} {2002})}\BibitemShut {NoStop}%
\bibitem [{\citenamefont {Quan}\ \emph {et~al.}(2006)\citenamefont {Quan},
  \citenamefont {Song}, \citenamefont {Liu}, \citenamefont {Zanardi},\ and\
  \citenamefont {Sun}}]{QSL06}%
  \BibitemOpen
  \bibfield  {author} {\bibinfo {author} {\bibfnamefont {H.}~\bibnamefont
  {Quan}}, \bibinfo {author} {\bibfnamefont {Z.}~\bibnamefont {Song}}, \bibinfo
  {author} {\bibfnamefont {X.}~\bibnamefont {Liu}}, \bibinfo {author}
  {\bibfnamefont {P.}~\bibnamefont {Zanardi}}, \ and\ \bibinfo {author}
  {\bibfnamefont {C.}~\bibnamefont {Sun}},\ }\href@noop {} {\bibfield
  {journal} {\bibinfo  {journal} {Phys. Rev. Lett.}\ }\textbf {\bibinfo
  {volume} {96}},\ \bibinfo {pages} {140604} (\bibinfo {year}
  {2006})}\BibitemShut {NoStop}%
\bibitem [{\citenamefont {Chaudhry}\ and\ \citenamefont {Gong}(2014)}]{CG14}%
  \BibitemOpen
  \bibfield  {author} {\bibinfo {author} {\bibfnamefont {A.~Z.}\ \bibnamefont
  {Chaudhry}}\ and\ \bibinfo {author} {\bibfnamefont {J.}~\bibnamefont
  {Gong}},\ }\href@noop {} {\bibfield  {journal} {\bibinfo  {journal} {Phys.
  Rev. A}\ }\textbf {\bibinfo {volume} {90}},\ \bibinfo {pages} {012101}
  (\bibinfo {year} {2014})}\BibitemShut {NoStop}%
\bibitem [{\citenamefont {Zurek}(1982)}]{Z82}%
  \BibitemOpen
  \bibfield  {author} {\bibinfo {author} {\bibfnamefont {W.~H.}\ \bibnamefont
  {Zurek}},\ }\href@noop {} {\bibfield  {journal} {\bibinfo  {journal} {Phys.
  Rev. D}\ }\textbf {\bibinfo {volume} {26}},\ \bibinfo {pages} {1862}
  (\bibinfo {year} {1982})}\BibitemShut {NoStop}%
\bibitem [{\citenamefont {Zeng}\ \emph {et~al.}(2011)\citenamefont {Zeng},
  \citenamefont {Tang}, \citenamefont {Zheng},\ and\ \citenamefont
  {Wang}}]{ZTZ11}%
  \BibitemOpen
  \bibfield  {author} {\bibinfo {author} {\bibfnamefont {H.-S.}\ \bibnamefont
  {Zeng}}, \bibinfo {author} {\bibfnamefont {N.}~\bibnamefont {Tang}}, \bibinfo
  {author} {\bibfnamefont {Y.-P.}\ \bibnamefont {Zheng}}, \ and\ \bibinfo
  {author} {\bibfnamefont {G.-Y.}\ \bibnamefont {Wang}},\ }\href@noop {}
  {\bibfield  {journal} {\bibinfo  {journal} {Phys. Rev. A}\ }\textbf {\bibinfo
  {volume} {84}},\ \bibinfo {pages} {032118} (\bibinfo {year}
  {2011})}\BibitemShut {NoStop}%
\bibitem [{\citenamefont {Haikka}\ \emph {et~al.}(2013)\citenamefont {Haikka},
  \citenamefont {Johnson},\ and\ \citenamefont {Maniscalco}}]{HJM13}%
  \BibitemOpen
  \bibfield  {author} {\bibinfo {author} {\bibfnamefont {P.}~\bibnamefont
  {Haikka}}, \bibinfo {author} {\bibfnamefont {T.}~\bibnamefont {Johnson}}, \
  and\ \bibinfo {author} {\bibfnamefont {S.}~\bibnamefont {Maniscalco}},\
  }\href@noop {} {\bibfield  {journal} {\bibinfo  {journal} {Phys. Rev. A}\
  }\textbf {\bibinfo {volume} {87}},\ \bibinfo {pages} {010103} (\bibinfo
  {year} {2013})}\BibitemShut {NoStop}%
\bibitem [{\citenamefont {Addis}\ \emph
  {et~al.}(2014{\natexlab{a}})\citenamefont {Addis}, \citenamefont {Brebner},
  \citenamefont {Haikka},\ and\ \citenamefont {Maniscalco}}]{ABH14}%
  \BibitemOpen
  \bibfield  {author} {\bibinfo {author} {\bibfnamefont {C.}~\bibnamefont
  {Addis}}, \bibinfo {author} {\bibfnamefont {G.}~\bibnamefont {Brebner}},
  \bibinfo {author} {\bibfnamefont {P.}~\bibnamefont {Haikka}}, \ and\ \bibinfo
  {author} {\bibfnamefont {S.}~\bibnamefont {Maniscalco}},\ }\href@noop {}
  {\bibfield  {journal} {\bibinfo  {journal} {Phys. Rev. A}\ }\textbf {\bibinfo
  {volume} {89}},\ \bibinfo {pages} {024101} (\bibinfo {year}
  {2014}{\natexlab{a}})}\BibitemShut {NoStop}%
\bibitem [{\citenamefont {Addis}\ \emph
  {et~al.}(2014{\natexlab{b}})\citenamefont {Addis}, \citenamefont {Bylicka},
  \citenamefont {Chru{\'s}ci{\'n}ski},\ and\ \citenamefont
  {Maniscalco}}]{ABC14}%
  \BibitemOpen
  \bibfield  {author} {\bibinfo {author} {\bibfnamefont {C.}~\bibnamefont
  {Addis}}, \bibinfo {author} {\bibfnamefont {B.}~\bibnamefont {Bylicka}},
  \bibinfo {author} {\bibfnamefont {D.}~\bibnamefont {Chru{\'s}ci{\'n}ski}}, \
  and\ \bibinfo {author} {\bibfnamefont {S.}~\bibnamefont {Maniscalco}},\
  }\href@noop {} {\bibfield  {journal} {\bibinfo  {journal} {Phys. Rev. A}\
  }\textbf {\bibinfo {volume} {90}},\ \bibinfo {pages} {052103} (\bibinfo
  {year} {2014}{\natexlab{b}})}\BibitemShut {NoStop}%
\bibitem [{\citenamefont {Ali}\ \emph {et~al.}(2015)\citenamefont {Ali},
  \citenamefont {Lo}, \citenamefont {Tu},\ and\ \citenamefont {Zhang}}]{ALT15}%
  \BibitemOpen
  \bibfield  {author} {\bibinfo {author} {\bibfnamefont {M.~M.}\ \bibnamefont
  {Ali}}, \bibinfo {author} {\bibfnamefont {P.-Y.}\ \bibnamefont {Lo}},
  \bibinfo {author} {\bibfnamefont {M.~W.-Y.}\ \bibnamefont {Tu}}, \ and\
  \bibinfo {author} {\bibfnamefont {W.-M.}\ \bibnamefont {Zhang}},\ }\href@noop
  {} {\bibfield  {journal} {\bibinfo  {journal} {Phys. Rev. A}\ }\textbf
  {\bibinfo {volume} {92}},\ \bibinfo {pages} {062306} (\bibinfo {year}
  {2015})}\BibitemShut {NoStop}%
\bibitem [{\citenamefont {Cucchietti}\ \emph {et~al.}(2005)\citenamefont
  {Cucchietti}, \citenamefont {Paz},\ and\ \citenamefont {Zurek}}]{CPZ05}%
  \BibitemOpen
  \bibfield  {author} {\bibinfo {author} {\bibfnamefont {F.}~\bibnamefont
  {Cucchietti}}, \bibinfo {author} {\bibfnamefont {J.~P.}\ \bibnamefont {Paz}},
  \ and\ \bibinfo {author} {\bibfnamefont {W.}~\bibnamefont {Zurek}},\
  }\href@noop {} {\bibfield  {journal} {\bibinfo  {journal} {Phys. Rev. A}\
  }\textbf {\bibinfo {volume} {72}},\ \bibinfo {pages} {052113} (\bibinfo
  {year} {2005})}\BibitemShut {NoStop}%
\bibitem [{\citenamefont {Rossini}\ \emph
  {et~al.}(2007{\natexlab{a}})\citenamefont {Rossini}, \citenamefont {Calarco},
  \citenamefont {Giovannetti}, \citenamefont {Montangero},\ and\ \citenamefont
  {Fazio}}]{RCG07}%
  \BibitemOpen
  \bibfield  {author} {\bibinfo {author} {\bibfnamefont {D.}~\bibnamefont
  {Rossini}}, \bibinfo {author} {\bibfnamefont {T.}~\bibnamefont {Calarco}},
  \bibinfo {author} {\bibfnamefont {V.}~\bibnamefont {Giovannetti}}, \bibinfo
  {author} {\bibfnamefont {S.}~\bibnamefont {Montangero}}, \ and\ \bibinfo
  {author} {\bibfnamefont {R.}~\bibnamefont {Fazio}},\ }\href@noop {}
  {\bibfield  {journal} {\bibinfo  {journal} {Phys. Rev. A}\ }\textbf {\bibinfo
  {volume} {75}},\ \bibinfo {pages} {032333} (\bibinfo {year}
  {2007}{\natexlab{a}})}\BibitemShut {NoStop}%
\bibitem [{\citenamefont {Rossini}\ \emph
  {et~al.}(2007{\natexlab{b}})\citenamefont {Rossini}, \citenamefont {Calarco},
  \citenamefont {Giovannetti}, \citenamefont {Montangero},\ and\ \citenamefont
  {Fazio}}]{RCG072}%
  \BibitemOpen
  \bibfield  {author} {\bibinfo {author} {\bibfnamefont {D.}~\bibnamefont
  {Rossini}}, \bibinfo {author} {\bibfnamefont {T.}~\bibnamefont {Calarco}},
  \bibinfo {author} {\bibfnamefont {V.}~\bibnamefont {Giovannetti}}, \bibinfo
  {author} {\bibfnamefont {S.}~\bibnamefont {Montangero}}, \ and\ \bibinfo
  {author} {\bibfnamefont {R.}~\bibnamefont {Fazio}},\ }\href@noop {}
  {\bibfield  {journal} {\bibinfo  {journal} {J. Phys. A: Math. Theor.}\
  }\textbf {\bibinfo {volume} {40}},\ \bibinfo {pages} {8033} (\bibinfo {year}
  {2007}{\natexlab{b}})}\BibitemShut {NoStop}%
\bibitem [{\citenamefont {Haikka}\ \emph {et~al.}(2012)\citenamefont {Haikka},
  \citenamefont {Goold}, \citenamefont {McEndoo}, \citenamefont {Plastina},\
  and\ \citenamefont {Maniscalco}}]{HGM12}%
  \BibitemOpen
  \bibfield  {author} {\bibinfo {author} {\bibfnamefont {P.}~\bibnamefont
  {Haikka}}, \bibinfo {author} {\bibfnamefont {J.}~\bibnamefont {Goold}},
  \bibinfo {author} {\bibfnamefont {S.}~\bibnamefont {McEndoo}}, \bibinfo
  {author} {\bibfnamefont {F.}~\bibnamefont {Plastina}}, \ and\ \bibinfo
  {author} {\bibfnamefont {S.}~\bibnamefont {Maniscalco}},\ }\href@noop {}
  {\bibfield  {journal} {\bibinfo  {journal} {Phys. Rev. A}\ }\textbf {\bibinfo
  {volume} {85}},\ \bibinfo {pages} {060101} (\bibinfo {year}
  {2012})}\BibitemShut {NoStop}%
\bibitem [{\citenamefont {Goan}\ \emph {et~al.}(2010)\citenamefont {Goan},
  \citenamefont {Jian},\ and\ \citenamefont {Chen}}]{GJC10}%
  \BibitemOpen
  \bibfield  {author} {\bibinfo {author} {\bibfnamefont {H.-S.}\ \bibnamefont
  {Goan}}, \bibinfo {author} {\bibfnamefont {C.-C.}\ \bibnamefont {Jian}}, \
  and\ \bibinfo {author} {\bibfnamefont {P.-W.}\ \bibnamefont {Chen}},\
  }\href@noop {} {\bibfield  {journal} {\bibinfo  {journal} {Phys. Rev. A}\
  }\textbf {\bibinfo {volume} {82}},\ \bibinfo {pages} {012111} (\bibinfo
  {year} {2010})}\BibitemShut {NoStop}%
\bibitem [{\citenamefont {Yang}\ and\ \citenamefont {Liu}(2008)}]{YL08}%
  \BibitemOpen
  \bibfield  {author} {\bibinfo {author} {\bibfnamefont {W.}~\bibnamefont
  {Yang}}\ and\ \bibinfo {author} {\bibfnamefont {R.-B.}\ \bibnamefont {Liu}},\
  }\href@noop {} {\bibfield  {journal} {\bibinfo  {journal} {Phys. Rev. Lett.}\
  }\textbf {\bibinfo {volume} {101}},\ \bibinfo {pages} {180403} (\bibinfo
  {year} {2008})}\BibitemShut {NoStop}%
\bibitem [{\citenamefont {Lee}\ \emph {et~al.}(2008)\citenamefont {Lee},
  \citenamefont {Witzel},\ and\ \citenamefont {Sarma}}]{LWS08}%
  \BibitemOpen
  \bibfield  {author} {\bibinfo {author} {\bibfnamefont {B.}~\bibnamefont
  {Lee}}, \bibinfo {author} {\bibfnamefont {W.}~\bibnamefont {Witzel}}, \ and\
  \bibinfo {author} {\bibfnamefont {S.~D.}\ \bibnamefont {Sarma}},\ }\href@noop
  {} {\bibfield  {journal} {\bibinfo  {journal} {Phys. Rev. Lett.}\ }\textbf
  {\bibinfo {volume} {100}},\ \bibinfo {pages} {160505} (\bibinfo {year}
  {2008})}\BibitemShut {NoStop}%
\bibitem [{\citenamefont {Chuang}\ and\ \citenamefont {Yamamoto}(1995)}]{CY95}%
  \BibitemOpen
  \bibfield  {author} {\bibinfo {author} {\bibfnamefont {I.~L.}\ \bibnamefont
  {Chuang}}\ and\ \bibinfo {author} {\bibfnamefont {Y.}~\bibnamefont
  {Yamamoto}},\ }\href {\doibase 10.1103/PhysRevA.52.3489} {\bibfield
  {journal} {\bibinfo  {journal} {Phys. Rev. A}\ }\textbf {\bibinfo {volume}
  {52}},\ \bibinfo {pages} {3489} (\bibinfo {year} {1995})}\BibitemShut
  {NoStop}%
\bibitem [{\citenamefont {Duan}\ \emph {et~al.}(2000)\citenamefont {Duan},
  \citenamefont {Giedke}, \citenamefont {Cirac},\ and\ \citenamefont
  {Zoller}}]{DGCZ00}%
  \BibitemOpen
  \bibfield  {author} {\bibinfo {author} {\bibfnamefont {L.-M.}\ \bibnamefont
  {Duan}}, \bibinfo {author} {\bibfnamefont {G.}~\bibnamefont {Giedke}},
  \bibinfo {author} {\bibfnamefont {J.~I.}\ \bibnamefont {Cirac}}, \ and\
  \bibinfo {author} {\bibfnamefont {P.}~\bibnamefont {Zoller}},\ }\href
  {\doibase 10.1103/PhysRevLett.84.4002} {\bibfield  {journal} {\bibinfo
  {journal} {Phys. Rev. Lett.}\ }\textbf {\bibinfo {volume} {84}},\ \bibinfo
  {pages} {4002} (\bibinfo {year} {2000})}\BibitemShut {NoStop}%
\bibitem [{\citenamefont {Paternostro}\ \emph {et~al.}(2003)\citenamefont
  {Paternostro}, \citenamefont {Kim},\ and\ \citenamefont {Ham}}]{PKH03}%
  \BibitemOpen
  \bibfield  {author} {\bibinfo {author} {\bibfnamefont {M.}~\bibnamefont
  {Paternostro}}, \bibinfo {author} {\bibfnamefont {M.~S.}\ \bibnamefont
  {Kim}}, \ and\ \bibinfo {author} {\bibfnamefont {B.~S.}\ \bibnamefont
  {Ham}},\ }\href {\doibase 10.1103/PhysRevA.67.023811} {\bibfield  {journal}
  {\bibinfo  {journal} {Phys. Rev. A}\ }\textbf {\bibinfo {volume} {67}},\
  \bibinfo {pages} {023811} (\bibinfo {year} {2003})}\BibitemShut {NoStop}%
\bibitem [{\citenamefont {Breuer}\ \emph {et~al.}(2009)\citenamefont {Breuer},
  \citenamefont {Laine},\ and\ \citenamefont {Piilo}}]{BLP09}%
  \BibitemOpen
  \bibfield  {author} {\bibinfo {author} {\bibfnamefont {H.-P.}\ \bibnamefont
  {Breuer}}, \bibinfo {author} {\bibfnamefont {E.-M.}\ \bibnamefont {Laine}}, \
  and\ \bibinfo {author} {\bibfnamefont {J.}~\bibnamefont {Piilo}},\ }\href
  {\doibase 10.1103/PhysRevLett.103.210401} {\bibfield  {journal} {\bibinfo
  {journal} {Phys. Rev. Lett.}\ }\textbf {\bibinfo {volume} {103}},\ \bibinfo
  {pages} {210401} (\bibinfo {year} {2009})}\BibitemShut {NoStop}%
\bibitem [{\citenamefont {Rivas}\ \emph {et~al.}(2010)\citenamefont {Rivas},
  \citenamefont {Huelga},\ and\ \citenamefont {Plenio}}]{RHP10}%
  \BibitemOpen
  \bibfield  {author} {\bibinfo {author} {\bibfnamefont {{\'A}.}~\bibnamefont
  {Rivas}}, \bibinfo {author} {\bibfnamefont {S.~F.}\ \bibnamefont {Huelga}}, \
  and\ \bibinfo {author} {\bibfnamefont {M.~B.}\ \bibnamefont {Plenio}},\
  }\href@noop {} {\bibfield  {journal} {\bibinfo  {journal} {Phys. Rev. Lett.}\
  }\textbf {\bibinfo {volume} {105}},\ \bibinfo {pages} {050403} (\bibinfo
  {year} {2010})}\BibitemShut {NoStop}%
\bibitem [{\citenamefont {Li}\ \emph {et~al.}(2017)\citenamefont {Li},
  \citenamefont {Hall},\ and\ \citenamefont {Wiseman}}]{LHW17}%
  \BibitemOpen
  \bibfield  {author} {\bibinfo {author} {\bibfnamefont {L.}~\bibnamefont
  {Li}}, \bibinfo {author} {\bibfnamefont {M.~J.}\ \bibnamefont {Hall}}, \ and\
  \bibinfo {author} {\bibfnamefont {H.~M.}\ \bibnamefont {Wiseman}},\
  }\href@noop {} {\bibfield  {journal} {\bibinfo  {journal} {arXiv preprint
  arXiv:1712.08879}\ } (\bibinfo {year} {2017})}\BibitemShut {NoStop}%
\bibitem [{\citenamefont {Hall}\ \emph {et~al.}(2014)\citenamefont {Hall},
  \citenamefont {Cresser}, \citenamefont {Li},\ and\ \citenamefont
  {Andersson}}]{MJL14}%
  \BibitemOpen
  \bibfield  {author} {\bibinfo {author} {\bibfnamefont {M.~J.~W.}\
  \bibnamefont {Hall}}, \bibinfo {author} {\bibfnamefont {J.~D.}\ \bibnamefont
  {Cresser}}, \bibinfo {author} {\bibfnamefont {L.}~\bibnamefont {Li}}, \ and\
  \bibinfo {author} {\bibfnamefont {E.}~\bibnamefont {Andersson}},\ }\href@noop
  {} {\bibfield  {journal} {\bibinfo  {journal} {Phys. Rev. A}\ }\textbf
  {\bibinfo {volume} {89}},\ \bibinfo {pages} {42120} (\bibinfo {year}
  {2014})}\BibitemShut {NoStop}%
\bibitem [{\citenamefont {Guarnieri}\ \emph {et~al.}(2014)\citenamefont
  {Guarnieri}, \citenamefont {Smirne},\ and\ \citenamefont {Vacchini}}]{GSV14}%
  \BibitemOpen
  \bibfield  {author} {\bibinfo {author} {\bibfnamefont {G.}~\bibnamefont
  {Guarnieri}}, \bibinfo {author} {\bibfnamefont {A.}~\bibnamefont {Smirne}}, \
  and\ \bibinfo {author} {\bibfnamefont {B.}~\bibnamefont {Vacchini}},\
  }\href@noop {} {\bibfield  {journal} {\bibinfo  {journal} {Phys. Rev. A}\
  }\textbf {\bibinfo {volume} {90}},\ \bibinfo {pages} {022110} (\bibinfo
  {year} {2014})}\BibitemShut {NoStop}%
\bibitem [{\citenamefont {Walls}\ and\ \citenamefont {Milburn}(1995)}]{WM07}%
  \BibitemOpen
  \bibfield  {author} {\bibinfo {author} {\bibfnamefont {D.~F.}\ \bibnamefont
  {Walls}}\ and\ \bibinfo {author} {\bibfnamefont {G.~J.}\ \bibnamefont
  {Milburn}},\ }\href@noop {} {\emph {\bibinfo {title} {Quantum Optics}}}\
  (\bibinfo  {publisher} {Springer-Verlag},\ \bibinfo {address} {Berlin},\
  \bibinfo {year} {1995})\BibitemShut {NoStop}%
\bibitem [{\citenamefont {Braunstein}\ and\ \citenamefont
  {Van~Loock}(2005)}]{BV05}%
  \BibitemOpen
  \bibfield  {author} {\bibinfo {author} {\bibfnamefont {S.~L.}\ \bibnamefont
  {Braunstein}}\ and\ \bibinfo {author} {\bibfnamefont {P.}~\bibnamefont
  {Van~Loock}},\ }\href@noop {} {\bibfield  {journal} {\bibinfo  {journal}
  {Rev. Mod. Phys.}\ }\textbf {\bibinfo {volume} {77}},\ \bibinfo {pages} {513}
  (\bibinfo {year} {2005})}\BibitemShut {NoStop}%
\bibitem [{\citenamefont {Breuer}\ and\ \citenamefont
  {Petruccione}(2007)}]{BP02}%
  \BibitemOpen
  \bibfield  {author} {\bibinfo {author} {\bibfnamefont {H.-P.}\ \bibnamefont
  {Breuer}}\ and\ \bibinfo {author} {\bibfnamefont {F.}~\bibnamefont
  {Petruccione}},\ }\href@noop {} {\emph {\bibinfo {title} {The Theory of Open
  Quantum Systems}}}\ (\bibinfo  {publisher} {Oxford University Press},\
  \bibinfo {address} {Oxford},\ \bibinfo {year} {2007})\BibitemShut {NoStop}%
\bibitem [{\citenamefont {Scully}\ and\ \citenamefont {Zubairy}(1997)}]{SZ99}%
  \BibitemOpen
  \bibfield  {author} {\bibinfo {author} {\bibfnamefont {M.~O.}\ \bibnamefont
  {Scully}}\ and\ \bibinfo {author} {\bibfnamefont {M.~S.}\ \bibnamefont
  {Zubairy}},\ }\href@noop {} {\emph {\bibinfo {title} {Quantum Optics}}}\
  (\bibinfo  {publisher} {Cambridge University Press},\ \bibinfo {address}
  {Cambridge},\ \bibinfo {year} {1997})\BibitemShut {NoStop}%
\bibitem [{Note2()}]{Note2}%
  \BibitemOpen
  \bibinfo {note} {The spectral density of the bath is often denoted as
  $J(\omega )=A \omega ^s {e}^{-\omega /\omega _c}$, where $A$ is a constant,
  ${e}^{-\omega /\omega _c}$ is the cutoff function and $\omega _c$ is the
  frequency cutoff~\cite {ZLX12}. Note that the coupling strength is also
  usually set as a constant. Therefore, the density of states can be written as
  $g_{\omega }=\omega ^s {e}^{-\omega /\omega _c}/g$. However, the cutoff is
  unnecessary in our model, so we choose the density of states in the form of
  Eq.~\protect \textup {\hbox {\mathsurround \z@ \protect \normalfont
  (\ignorespaces \ref {eq-gomega}\unskip \@@italiccorr )}} \protect \color
  {black}}\BibitemShut {NoStop}%
\bibitem [{\citenamefont {Olver}(2010)}]{O10}%
  \BibitemOpen
  \bibfield  {author} {\bibinfo {author} {\bibfnamefont {F.~W.}\ \bibnamefont
  {Olver}},\ }\href@noop {} {\emph {\bibinfo {title} {NIST Handbook of
  Mathematical Functions Hardback and CD-ROM}}}\ (\bibinfo  {publisher}
  {Cambridge University Press},\ \bibinfo {address} {Cambridge},\ \bibinfo
  {year} {2010})\BibitemShut {NoStop}%
\bibitem [{\citenamefont {Rivas}\ \emph {et~al.}(2014)\citenamefont {Rivas},
  \citenamefont {Huelga},\ and\ \citenamefont {Plenio}}]{RHP14}%
  \BibitemOpen
  \bibfield  {author} {\bibinfo {author} {\bibfnamefont {{\'A}.}~\bibnamefont
  {Rivas}}, \bibinfo {author} {\bibfnamefont {S.~F.}\ \bibnamefont {Huelga}}, \
  and\ \bibinfo {author} {\bibfnamefont {M.~B.}\ \bibnamefont {Plenio}},\
  }\href@noop {} {\bibfield  {journal} {\bibinfo  {journal} {Rep. Prog. Phys.}\
  }\textbf {\bibinfo {volume} {77}},\ \bibinfo {pages} {094001} (\bibinfo
  {year} {2014})}\BibitemShut {NoStop}%
\bibitem [{\citenamefont {Breuer}\ \emph {et~al.}(2016)\citenamefont {Breuer},
  \citenamefont {Laine}, \citenamefont {Piilo},\ and\ \citenamefont
  {Vacchini}}]{BLP16}%
  \BibitemOpen
  \bibfield  {author} {\bibinfo {author} {\bibfnamefont {H.-P.}\ \bibnamefont
  {Breuer}}, \bibinfo {author} {\bibfnamefont {E.-M.}\ \bibnamefont {Laine}},
  \bibinfo {author} {\bibfnamefont {J.}~\bibnamefont {Piilo}}, \ and\ \bibinfo
  {author} {\bibfnamefont {B.}~\bibnamefont {Vacchini}},\ }\href@noop {}
  {\bibfield  {journal} {\bibinfo  {journal} {Rev. Mod. Phys.}\ }\textbf
  {\bibinfo {volume} {88}},\ \bibinfo {pages} {021002} (\bibinfo {year}
  {2016})}\BibitemShut {NoStop}%
\bibitem [{Note1()}]{Note1}%
  \BibitemOpen
  \bibinfo {note} {Manuscript in preparation.}\BibitemShut {Stop}%
\bibitem [{\citenamefont {Axler}(1997)}]{A97}%
  \BibitemOpen
  \bibfield  {author} {\bibinfo {author} {\bibfnamefont {S.~J.}\ \bibnamefont
  {Axler}},\ }\href@noop {} {\emph {\bibinfo {title} {Linear Algebra Done
  Right}}},\ Vol.~\bibinfo {volume} {2}\ (\bibinfo  {publisher} {Springer},\
  \bibinfo {address} {New York},\ \bibinfo {year} {1997})\BibitemShut {NoStop}%
\bibitem [{\citenamefont {Carmichael}(1999)}]{HC991}%
  \BibitemOpen
  \bibfield  {author} {\bibinfo {author} {\bibfnamefont {H.~J.}\ \bibnamefont
  {Carmichael}},\ }\href@noop {} {\emph {\bibinfo {title} {{Statistical Methods
  in Quantum Optics 1, Master Equations and Fokker-Planck Equations}}}}\
  (\bibinfo  {publisher} {Springer},\ \bibinfo {address} {Berlin},\ \bibinfo
  {year} {1999})\BibitemShut {NoStop}%
\bibitem [{\citenamefont {Kenfack}\ and\ \citenamefont
  {{\.Z}yczkowski}(2004)}]{KZ04}%
  \BibitemOpen
  \bibfield  {author} {\bibinfo {author} {\bibfnamefont {A.}~\bibnamefont
  {Kenfack}}\ and\ \bibinfo {author} {\bibfnamefont {K.}~\bibnamefont
  {{\.Z}yczkowski}},\ }\href
  {http://iopscience.iop.org/article/10.1088/1464-4266/6/10/003/meta}
  {\bibfield  {journal} {\bibinfo  {journal} {J. Opt. B Quantum Semiclassical
  Opt}\ }\textbf {\bibinfo {volume} {6}},\ \bibinfo {pages} {396} (\bibinfo
  {year} {2004})}\BibitemShut {NoStop}%
\bibitem [{\citenamefont {Serafini}\ \emph {et~al.}(2005)\citenamefont
  {Serafini}, \citenamefont {Paris}, \citenamefont {Illuminati},\ and\
  \citenamefont {De~Siena}}]{SPID05}%
  \BibitemOpen
  \bibfield  {author} {\bibinfo {author} {\bibfnamefont {A.}~\bibnamefont
  {Serafini}}, \bibinfo {author} {\bibfnamefont {M.}~\bibnamefont {Paris}},
  \bibinfo {author} {\bibfnamefont {F.}~\bibnamefont {Illuminati}}, \ and\
  \bibinfo {author} {\bibfnamefont {S.}~\bibnamefont {De~Siena}},\ }\href
  {http://iopscience.iop.org/article/10.1088/1464-4266/7/4/R01/meta} {\bibfield
   {journal} {\bibinfo  {journal} {J. Opt. B: Quantum Semiclassical Opt.}\
  }\textbf {\bibinfo {volume} {7}},\ \bibinfo {pages} {R19} (\bibinfo {year}
  {2005})}\BibitemShut {NoStop}%
\bibitem [{\citenamefont {Mahan}(2000)}]{M13}%
  \BibitemOpen
  \bibfield  {author} {\bibinfo {author} {\bibfnamefont {G.~D.}\ \bibnamefont
  {Mahan}},\ }\href@noop {} {\emph {\bibinfo {title} {Many-Particle Physics}}}\
  (\bibinfo  {publisher} {Springer},\ \bibinfo {address} {New York},\ \bibinfo
  {year} {2000})\BibitemShut {NoStop}%
\bibitem [{\citenamefont {Zhang}\ \emph {et~al.}(2012)\citenamefont {Zhang},
  \citenamefont {Lo}, \citenamefont {Xiong}, \citenamefont {Tu},\ and\
  \citenamefont {Nori}}]{ZLX12}%
  \BibitemOpen
  \bibfield  {author} {\bibinfo {author} {\bibfnamefont {W.-M.}\ \bibnamefont
  {Zhang}}, \bibinfo {author} {\bibfnamefont {P.-Y.}\ \bibnamefont {Lo}},
  \bibinfo {author} {\bibfnamefont {H.-N.}\ \bibnamefont {Xiong}}, \bibinfo
  {author} {\bibfnamefont {M.~W.-Y.}\ \bibnamefont {Tu}}, \ and\ \bibinfo
  {author} {\bibfnamefont {F.}~\bibnamefont {Nori}},\ }\href {\doibase
  10.1103/PhysRevLett.109.170402} {\bibfield  {journal} {\bibinfo  {journal}
  {Phys. Rev. Lett.}\ }\textbf {\bibinfo {volume} {109}},\ \bibinfo {pages}
  {170402} (\bibinfo {year} {2012})}\BibitemShut {NoStop}%
\end{thebibliography}%
	
\end{document}